\documentclass[iop]{emulateapj}
\usepackage{amsmath} 
\usepackage{aas_macros,amssymb}
\usepackage{natbib,graphicx,layouts}
\usepackage{color,hyperref}
\definecolor{linkcolor}{rgb}{0,0,0.5}
\definecolor{notecolor}{rgb}{0.8,0,0}
\hypersetup{colorlinks=true,linkcolor=linkcolor,citecolor=linkcolor,
  filecolor=linkcolor,urlcolor=linkcolor}

\shorttitle{DLAs and Population~III stars}
\shortauthors{Kulkarni et al.}

\begin{document}

\title{Chemical enrichment of Damped Lyman Alpha
  systems as a direct constraint on Population III star formation}
\author{Girish Kulkarni\altaffilmark{1}, 
  Emmanuel Rollinde\altaffilmark{2}, 
  Joseph F.~Hennawi\altaffilmark{1}, 
  Elisabeth Vangioni\altaffilmark{2}} 
\altaffiltext{1}{Max Planck Institute for Astronomy, K\"onigstuhl 17,
  69117 Heidelberg, Germany;
  \email{girish@mpia-hd.mpg.de}}
\altaffiltext{2}{Institut d'Astrophysique de Paris, UMR 7095, UPMC,
  Paris VI, 98 bis boulevard Arago, 75014 Paris, France}

\begin{abstract}
Observations of damped Ly$\alpha$ absorbers (DLAs) can be used to
measure gas-phase metallicities at large cosmological lookback times
with high precision. Furthermore, relative abundances can still be
measured accurately deep into the reionization epoch ($z > 6$) using
transitions redward of Ly$\alpha$, even though Gunn-Peterson
absorption precludes measurement of neutral hydrogen. In this paper we
study the chemical evolution of DLAs using a model for the coupled
evolution of galaxies and the intergalactic medium (IGM), which is
constrained by a variety of observations. Our goal is to explore the
influence of Population~III stars on the abundance patterns of DLAs to
determine the degree to which abundance measurements can discriminate
between different Population~III stellar initial mass functions
(IMFs). We include effects such as inflows onto galaxies due to
cosmological accretion and outflows from galaxies due to supernova
feedback. A distinct feature of our model is that it self-consistently
calculates the effect of Population III star formation on the
reionization of an inhomogeneous IGM, thus allowing us to calculate
the thermal evolution of the IGM and implement photoionization
feedback on low-mass galaxy formation.  We find that if the critical
metallicity of Population~III to II/I transition is $\lesssim 10^{-4}
Z_{\odot}$, then the cosmic Population III star formation rate (SFR)
drops to zero for $z<8$. Nevertheless, at high redshift ($z\sim 6$)
chemical signatures of Population~III stars remain in low mass
galaxies (halo mass $\lesssim 10^9$ M$_{\odot}$). This is because
photoionization feedback suppresses star formation in these galaxies
until relatively low redshift ($z\sim 10$), and the chemical record of
their initial generation of Population~III stars is retained.  We
model DLAs as these low-mass galaxies, and assign to them a
mass-dependent H~\textsc{i} absorption cross-section in order to
predict the expected distribution of DLA abundance ratios.  We find
that these distributions are anchored towards abundance ratios set by
Population~II supernovae yields, but they exhibit a tail which depends
significantly on the Population~III IMF for $z > 5$. Thus, a sample of
DLA metallicity and relative abundance measurements at high-redshift
holds the promise to constrain Population~III enrichment and the
Population~III IMF.  We find that sample of just 10 DLAs with relative
abundances measured to an accuracy of 0.1 dex is sufficient to
constrain the Population~III IMF at 4$\sigma$.  These constraints may
prove stronger than other probes of Population~III enrichment, such as
metal-poor stars and individual metal-poor DLAs. Our results provide a
global picture of the thermal, ionization, and chemical evolution of
the Universe, and have the potential to rule out certain
Population~III scenarios.
\end{abstract}


\section{Introduction}
\label{s:introduction}

The first stars in the Universe formed out of the primordial
interstellar media (ISM) of the first galaxies and are therefore
expected to be metal-free.  Formation of these so-called
Population~III stars is an important milestone in cosmic evolution as
the UV photons produced by these objects presumably initiated the
process of reionization of the intergalactic medium (IGM).
Reionization, and the consequent reheating of the IGM result in new
physical processes, such as photoionization feedback, that are
expected to influence subsequent galaxy formation and evolution.
Apart from reionization, Population~III stars are also expected to
contribute to chemical feedback.  The metal yield of these stars can
pollute the ISM to create conditions for formation of metal-enriched
Population~II stars that we observe today.  Population~III stars have
also been invoked to seed the formation of supermassive black holes
seen in centres of galaxies at low redshift.  In the $\Lambda$CDM
universe, in which galaxies begin their evolution with primordial
abundances, Population~III must have formed at some point.  Still,
their observational signatures and evidence for their existence is
still elusive.  (The only hints so far are from extremely metal-poor
stars in the Milky Way's halo, as we describe below.)  These stars
thus form a significant piece of the puzzle of galaxy formation
\citep{2004ARA&A..42...79B, 2010hdfs.book.....L}.

The initial conditions of galaxy formation are well known thanks to
cosmological constraints at the last scattering surface
\citep{2005SSRv..116..625C, 2011ARA&A..49..373B}.  In the first
galaxies that result from these initial conditions (at $z\gtrsim 20$
in dark matter haloes with mass of about $10^6$ M$_\odot$), gas can
fragment to form Population~III stars.  The chemistry of gas in these
haloes is simple, and formation of the first generation of stars is
unaffected by complicated effects like feedback.  Physical factors
such as magnetic fields are expected to play a negligible role in the
formation of Population~III stars as dynamically significant magnetic
fields are not expected to be present in Population~III star-forming
clouds \citep{2004ARA&A..42...79B, 2012ApJ...745..154T}.  We can
therefore hope that the physics of Population~III star formation is
simpler than that of star formation in the present day universe.

Nonetheless, properties of Population~III stars, such as their mass,
are not well-understood.  This is primarily because of the wide range
of scales involved in the fragmentation and collapse of gas in high
redshift haloes.  On the galactic scale, the mass of the first
galaxies is debatable \citep{2011ARA&A..49..373B}.  Since the
molecular hydrogen cooling threshold provides a lower bound on halo
mass for gas cooling and fragmentation, we expect that the first stars
will be hosted by haloes with virial temperature $T_\mathrm{vir}=10^3$
K.  However, molecular hydrogen is fragile and can be easily destroyed
as the IGM is transparent at the required energies.  If the molecular
hydrogen is destroyed by the Lyman-Werner background produced by the
very first few stars, then the formation of new stars could be delayed
up to a lower redshift in all but the most massive haloes
\citep[e.g.,][]{2001PhR...349..125B}.  On the proto-stellar scale, it
is unclear whether the gas collected in these haloes will fragment, or
what will be the mass of the fragments if it does.  It has been
suggested that dynamical effects can cause fragmentation even when
metal-line cooling is not available \citep{2012arXiv1203.6842D}.  But
the magneto-hydrodynamics in this regime is very difficult to
simulate.  In summary, both the initial mass function (IMF) as well as
the cosmic star formation rate (SFR) history of Population~III stars
remain unclear.  We are thus forced to ask if there are any empirical
constraints on these quantities.

Unfortunately, no understanding of Population~III stars has emerged
from the observational side either.  Currently, two lines of
observational inquiry have aimed at an unambiguous detection of the
chemical signatures of Population~III stars.  These probe either (1)
low-mass extremely metal-poor stars in the Galaxy's halo, or (2)
metal-poor neutral gas reservoirs, presumably galaxies, seen in quasar
absorption spectra at high redshift ($z\gtrsim 2$) as damped
Ly$\alpha$ absorbers (DLAs).  In the first case, Galactic stellar
archaeology has been considered as an important probe of high redshift
star formation environment \citep{2011AAS...21732901F}.  These
long-lived stars could potentially carry signatures of enrichment by
Population~III supernovae \citep{2009MNRAS.398.1782R,
  2013arXiv1303.1791S}.  However, no detection of a truly metal-free
star has been made to date.  The lowest-metallicity star known has
[Fe/H] $=-4.88\pm 0.12$ \citep{2011Natur.477...67C}, although current
observations have a potential sensitivity of metallicity down to
[Fe/H] $=-9.8$ \citep{2011arXiv1102.1748F}.  Moreover, although they
are a valuable tool for constraining Galactic chemical evolution, it
is not clear if metal-poor stars are an unambiguous probe of the
cosmic SFR and IMF of Population~III stars, because these observations
are limited to the peculiar star formation and assembly history of the
Galaxy and because abundance ratios in a single metal-poor star cannot
probe the whole Population~III IMF.  In the second case, metal-poor
DLAs at $z\sim 2$ likely represent systems that form in peculiar,
pristine patches of the IGM.  They could be direct descendants of high
redshift haloes and, due to inefficient star formation, probably
retain signature of Population~III star formation
\citep{2009MNRAS.395L...6S, 2012MNRAS.421L..29S}.  However,
observations of such systems have also failed to show up chemical
abundance patterns that cannot be explained with Population~II yields
\citep{2005MNRAS.360..447M, 2011MNRAS.417.1534C, 2012ApJ...757L..22F}.

In the absence of current observational constraints, there are
suggestions that future facilities such as the James Webb Space
Telescope (JWST) and the Large Synoptic Survey Telescope (LSST) may
detect pair-instability supernovae (PISN) from any high-mass
Population~III stars in situ at high redshift
\citep{2012MNRAS.422.2701P, 2012MNRAS.423.2203P}.  It has also been
argued that massive Population~III stars could trigger long GRBs
\citep{2006ApJ...642..382B}.  Detection of these objects will also
potentially probe the first stars \citep{2012grb..confE.137L}.

In this paper, we study the possibility of probing the Population~III
SFR and IMF by measuring chemical abundance ratios in low mass
galaxies at high redshift ($z\sim 6$).  We set up a conservative model
of evolution of galaxies and the IGM, constrained by low-redshift
observations, and show that different Population~III IMFs leave
distinct signatures in the metal abundance ratios of low mass galaxies
at high redshift.  Further, these galaxies can be observed as DLAs in
the spectra of any background quasars.  These absorption-line
observations can be used to measure gas-phase metallicities in these
galaxies with high precision.  Even deep into the reionization epoch
($z > 6$), abundance ratios can still be measured accurately in such
systems using transitions redward of Ly$\alpha$.  We model DLAs as
these low-mass galaxies by assigning them a mass-dependent
H~\textsc{i} absorption cross section, and predict the expected
distribution of DLA abundance ratios at various redshifts. We find
that at high redshift \emph{the distribution of DLAs in the
  abundance-ratio space can be mapped to a combination of
  Population~III SFR and IMF.}  This can be a powerful probe of the
first stars.

We describe our modeling of the evolution of galaxies and the IGM in
Section \ref{sec:model}. Our results are presented in Section
\ref{sec:results}.  We discuss some caveats in Section
\ref{sec:discuss} and present a summary of conclusions in Section
\ref{sec:conclude}.  Throughout this paper, we assume a WMAP 9-yr
cosmology \citep{2012arXiv1212.5226H}. 

\section{Modeling galaxy and IGM evolution}
\label{sec:model}

Given a dark matter halo of mass $M_h$ at $z=0$, we can calculate its
complete \emph{mean} assembly history.  This can be done using the
extended Press-Schechter formalism, or using large volume N-body
simulations of structure formation.  We prefer the latter approach for
its accuracy, and use fitting functions provided by
\citet{2010MNRAS.406.2267F}.  This lets us calculate the mean dark
matter mass $M$ of a halo at any stage of its assembly history.  We
use this to set up a model of baryonic evolution inside the halo.  In
simplest terms, this model assumes that a halo (1) accretes baryons
through cosmological accretion, (2) forms stars from any gas contained
in the halo for sufficiently long duration, and (3) ejects baryons due
to supernova-induced outflows.

With these assumptions, we can associate with each dark matter halo a
stellar mass $M_*$ and a gas mass $M_g$ at any point in its assembly
history.  Further, using stellar population synthesis, we can also
calculate the total amount of metals $M_Z$ and the rate of UV photon
production $\dot n_\mathrm{ph}$ in a halo at any stage of its assembly
history.  All of these quantities are functions of the redshift-zero
halo mass ($M_h$).  We repeat this exercise for a number of
logarithmically spaced halo masses between $M_h^\mathrm{low}$ and
$M_h^\mathrm{high}$, two values that are chosen using a convergence
criterion.  The integrated UV photon yield of all halos in this mass
range is then used in a reionization model to calculate the ionization
and thermal history of the IGM.  Similarly, the integrated outflow
yield is used with a chemical enrichment model to calculate the
average IGM metallicity.  The IGM ionization, thermal, and chemical
evolution models feed back into the galaxy formation model through
photoionization feedback, which controls the thermal cut-off on the
mass of star-forming haloes, and by changing the metallicity of
inflows from the IGM to the galaxies.

This galaxy formation picture is often referred to as the ``bathtub''
model, and has been used to study various problems in galaxy formation
to a reasonable degree of accuracy \citep{2010ApJ...718.1001B,
  2012ApJ...753...16K}.  We will now describe our method in detail.  A
summary of free parameters of our model is presented in Table
\ref{table:params}.

\subsection{Halo growth}

The mean mass growth rate of dark matter haloes in the $\Lambda$CDM
cosmology can be parameterised by the fitting function
\citep{2010MNRAS.406.2267F}
\begin{multline}
\dot M(z) = 46.1\,\frac{\mathrm{M}_\odot}{\mathrm{yr}}\,(1+1.1z)\\\times\sqrt{\Omega_m(1+z)^3+\Omega_\Lambda}\,\left(\frac{M(z)}{10^{12}\mathrm{M}_\odot}\right)^{1.1},
\label{eqn:halo_assembly}
\end{multline}
where $M(z)$ denotes the mean mass in the main branch of the merger
tree of a halo at redshift $z$. This growth rate corresponds to the
mean rate of growth of halo mass.  This relation can be understood in
terms of the extended Press-Schechter picture of structure formation
\citep{2008MNRAS.383..615N}.  By comparing it with cosmological
simulations, \citet{2010MNRAS.406.2267F} have shown that it is
accurate over a mass range spanning at least five orders of magnitude
($10^{10}$ M$_\odot$ to $10^{15}$ M$_\odot$ at $z=0$) and a wide range
of redshifts ($0\leq z\lesssim 15$).  Integrating Equation
(\ref{eqn:halo_assembly}) with the boundary condition $M(z=0)=M_h$
gives the mean mass of a halo at any redshift.  Note that this
describes the evolution of the dark matter mass of the halo.

\subsection{Baryons} 

Baryonic content of galaxies is influenced by gas inflows, gas
outflows, and star formation.  Hydrodynamical simulations suggest that
these process tend to be in equilibrium \citep{2012MNRAS.421...98D}.
We now describe these processes and calculate the mean evolution of
gas mass ($M_g$), stellar mass ($M_*$), and metal mass ($M_Z$) in each
halo mass bin.  The metal mass term describes the mass of the metals
in the gas phase (ISM); it excludes the metals that are locked up
inside stars.

\subsubsection{Gas}

Gas inflows replenish the gas content of haloes as they evolve.  We
assume that the mean gas inflow rate is proportional to the mean halo
growth rate
\begin{equation}
  \dot M_{g,\mathrm{in}}(z) = f_{g,\mathrm{in}}\left(\frac{\Omega_b}{\Omega_m}\right)\dot M(z).
\end{equation}
We follow \citet{2010ApJ...718.1001B} and take the constant of
proportionality to be $f_{g,\mathrm{in}}=0.7$ , except when the halo
mass is below the filtering mass $M_\mathrm{min}$ (described below),
in which case there is no gas inflow and we set $f_{g,\mathrm{in}}=0$.
The choice of these values is motivated by the comparison of measured
SFR in massive high-redshift galaxies with predicted gas accretion
rates \citep{2008ApJ...688..789G, 2009Natur.457..451D,
  2010ApJ...717..323B}.

Once inside the halo, some of the accreted gas is lost to star
formation.  The amount of gas lost is
\begin{multline}
  \dot M_{g, \mathrm{sf}}(z) = -\psi(z)\\+\int_{m_l}^{m_u}dm\,\phi(m)\cdot\psi[t(z)-\tau(m)]\cdot[m-m_r(m)], 
  \label{eq:gassf}
\end{multline}
where the $\psi$ is the star formation rate (which we describe below),
and the second term accounts for mass loss from evolving stars via
stellar winds and supernova explosions.  In the second term, $\phi(m)$
is the stellar IMF, $m_r(m)$ is the remnant mass left by a star with
initial mass $m$, $\tau(m)$ is its lifetime, and $t(z)$ is the cosmic
time.  The lower limit $m_l$ of the integral is such that
$\tau(m_l)=t(z)$.  (The upper limit $m_u$ corresponds to the high-mass
limit of the stellar IMF, which we discuss below.)  The instantaneous
recycling approximation (IRA) is often used in calculating the mass
loss from stars \citep[e.g.,][]{2012ApJ...753...16K}.  In this
approximation, stars with initial mass above a certain value $m_0$ are
assumed to die instantaneously ($\tau=0$) while those with initial
mass less than $m_0$ are assumed to live forever ($\tau=\infty$).
However, although this is a reasonable assumption to make while
studying, say, the solar neighbourhood, it is not a good approximation
at high redshift, when stellar lifetimes are comparable to the Hubble
time.  Therefore, we do \emph{not} use IRA in
Equation~(\ref{eq:gassf}).

Gas is also lost due to outflows resulting out of supernova-driven
winds.  This effect is expected to be proportional to the star
formation rate and inversely proportional to the depth of the halo
potential well \citep{2006ApJ...647..773D, 2008ApJ...674..151E,
  2010ApJ...718.1001B, 2012ApJ...753...16K}.  We write the mass loss
due to outflows as
\begin{multline}
  \dot M_{g, \mathrm{out}}(z) = -
  \frac{2\epsilon}{v^2_\mathrm{esc}}\int_{m_l}^{m_u}dm\,\left\{\phi(m)\right.\\\times\left.\psi[t(z)-\tau(m)]\cdot  E_\mathrm{kin}(m)\right\},
\end{multline}
where $v^2_\mathrm{esc}=2GM/R_\mathrm{vir}$ is the escape velocity of
the halo, and $E_\mathrm{kin}(m)$ is the kinetic energy released by a
star of mass $m$.  The parameter $\epsilon$ is fixed by calculating
the total baryon fraction in structures, $f_{b,\mathrm{struct}}$, at
$z=0$ \citep{2004ApJ...616..643F}.

Combining the contribution of above three processes of star formation,
inflows, and outflows, the evolution of the gas mass of a halo can be
written as
\begin{equation}
  \dot M_g(z) = \dot M_{g,\mathrm{in}}(z) + \dot M_{g, \mathrm{sf}}(z) + \dot M_{g, \mathrm{out}}(z),
\end{equation}
where each term is redshift- and halo-mass-dependent.

\subsubsection{Stars}

We assume that the star formation rate, $\psi$, in a halo tracks the
total amount of cold gas, $M_\mathrm{cool}$, inside that halo.  Thus
\begin{equation}
  \psi = f_*\left(\frac{M_\mathrm{cool}}{t_\mathrm{dyn}}\right),
  \label{eqn:sf}
\end{equation}
where $t_\mathrm{dyn}$ is the halo dynamical time and $f_*$ is a free
parameter.  The cold gas mass is calculated by defining a local
cooling rate within the halo,
\begin{equation}
  t_\mathrm{cool}(r)=\frac{3k_BT\rho_g(r)}{2\mu n_H^2(r)\Lambda(T)},
\end{equation}
where $\rho_g$ is the (spherically symmetric) gas density profile,
$n_H$ is the hydrogen number density, $\mu$ is the molecular weight
and $\Lambda$ is the cooling function.  A cooling radius
$r_\mathrm{cool}$ can then be defined by
\begin{equation}
  t_\mathrm{cool}(r_\mathrm{cool})=t_\mathrm{dyn},
\end{equation}
which gives the cooling rate as
\begin{equation}
  \frac{dM_\mathrm{cool}}{dt} = 4\pi\rho_g(r_\mathrm{cool})r^2_\mathrm{cool}\frac{dr_\mathrm{cool}}{dt}.
  \label{eqn:mcool}
\end{equation}
Initially all gas entering the halo is shock-heated to the virial
temperature.  Integrating Equation~(\ref{eqn:mcool}) then gives the
amount of cool gas at any time.  We use a gas profile close to
isothermal, which fits the results of \citet{2003MNRAS.339..312S} to
better than 10\%.  Equation (\ref{eqn:sf}) follows from the empirical
Kennicutt-Schmidt (KS) relation \citep{2007ApJ...669..289K} in the
limit of marginally unstable disks.  The KS relation does not exhibit
any evolution till $z\sim 2$ \citep{2010ApJ...714L.118D}.

The galaxy dynamical time can be written as
\begin{equation}
  t_\mathrm{dyn} = 2\times 10^7 \mathrm{yr} \left(\frac{R_{1/2}}{4\ \mathrm{kpc}}\right)\left(\frac{V_c}{200\ \mathrm{km}\,\mathrm{s}^{-1}}\right)^{-1},
\end{equation}
where $R_{1/2}$ is galaxy disk scale-length and $V_c$ is the halo
circular velocity \citep{2010ApJ...718.1001B}.  We take the disk
scale-length to be proportional to the halo virial radius
\begin{equation}
  R_{1/2} = 0.05 \left(\frac{\lambda}{0.1}\right)R_\mathrm{vir},
\end{equation}
where the halo spin parameter $\lambda$ is taken to be $0.07$
\citep{2012ApJ...753...16K}.  The halo circular velocity is given by
\begin{multline}
V_\mathrm{c}=23.4\left(\frac{M}{10^8h^{-1}M_\odot}\right)^{1/3}\left[\frac{\Omega_m}{\Omega_m^z}\frac{\Delta_c}{18\pi^2}\right]^{1/6}\\\times\left(\frac{1+z}{10}\right)^{1/2} \mathrm{km}/\mathrm{s},
\end{multline}
where 
\begin{equation}
\Omega_m^z=\frac{\Omega_m(1+z)^3}{\Omega_m(1+z)^3+\Omega_\Lambda+\Omega_k(1+z)^2},
\end{equation}
and $\Delta_c$ is the halo overdensity relative to the cosmological
critical density, given by 
\begin{equation}
\Delta_c=18\pi^2+82d-39d^2,
\end{equation}
for the $\Lambda$CDM cosmology, where $d=\Omega_m^z-1$
\citep{2001PhR...349..125B}.  The virial radius is given by
\begin{multline}
  R_\mathrm{vir}=0.784\left(\frac{M}{10^8 h^{-1}M_\odot}\right)^{1/3}\left[\frac{\Omega_m}{\Omega_m^z}\frac{\Delta_c}{18\pi^2}\right]^{-1/3}\\\times\left(\frac{1+z}{10}\right)^{-1} h^{-1}\,\mathrm{kpc}.
\end{multline}

The total stellar mass in the halo evolves as
\begin{multline}
  \dot M_*(z) = - \dot M_{g, \mathrm{sf}}(z) = 
  \psi(z)\\-\int_{m_l}^{m_u}dm\,\phi(m)\cdot\psi[t(z)-\tau(m)]\cdot[m-m_r(m)],
  \label{eqn:halosfr}
\end{multline}
where, as in Equation (\ref{eq:gassf}), the second term on the right
hand side accounts for mass loss from stars.

\subsubsection{Metals}

The initial mass of metals in a halo is zero.  Metals are produced in
stars in the halo and are mixed in the halo gas when stars explode as
supernovae, and due to stellar winds.  The total mass of metals $M_Z$
in the halo is also affected by inflows from the IGM and outflows into
it. The inflow term is given by
\begin{equation}
  \dot M_{Z,\mathrm{in}}(z) = Z_\mathrm{IGM}(z)\dot M_{g,\mathrm{in}},
\end{equation}
where $Z_\mathrm{IGM}$ is the metal abundance in the IGM, which we
describe below.

Metal outflows can be described in similar fashion, as
\begin{equation}
  \dot M_{Z,\mathrm{out}}(z) = -Z_g(z)\dot M_{g,\mathrm{out}},
\end{equation}
where $Z_g\equiv M_Z/M_g$ is the gas metal abundance in the halo.

Lastly, the effect of star formation can be expressed as
\citep{2012ApJ...753...16K}
\begin{multline}
   \dot M_{Z,\mathrm{sf}}(z) = \int_{m_l}^{m_u}dm\,\phi(m)\cdot\psi[t(z)-\tau(m)]\\\times mp_Z(m)(1-\zeta).
\label{eqn:sfrz}
\end{multline}
Here, $p_Z(m)$ is the mass fraction of a star of initial mass $m$ that
is converted to metals and ejected.  The factor $1-\zeta$ takes into
account the fact that not all of the newly enriched material is going
to mix in the ISM.  Some of it will be directly ejected out of the
halo due to supernova explosions.  We follow
\citet{2012ApJ...753...16K} and write
\begin{equation}
  \zeta = \zeta_l\,\exp{(-M_{h,12}/M_\mathrm{ret})},
  \label{eqn:zeta} 
\end{equation}
where $M_{h,12}$ is the halo mass in units of $10^{12}$ M$_\odot$ and
we set the parameters $M_\mathrm{ret}=0.3$ and $\zeta_l=0.9$.  These
values are expected to depend on the stellar IMF and the problem
geometry. Our choice of these values is motivated by the simulations
of \citet{1999ApJ...513..142M} and the findings of
\citet{2012ApJ...753...16K}.  These authors find that changing these
quantities within reasonable limits has only a modest effect on the
star formation and metallicity evolution.  

The evolution of the metal mass of a halo can now be written as
\begin{equation}
  \dot M_Z(z) = \dot M_{Z,\mathrm{sf}}(z) + \dot M_{Z,\mathrm{out}}(z) +  \dot M_{Z,\mathrm{in}}(z).
\end{equation}

\subsection{IMF}

\begin{deluxetable}{lll}
\tablecaption{Stellar IMFs used in this paper\label{table:models}}
\tablehead{\colhead{Model} & \colhead{Pop.~III IMF} & \colhead{Pop.~II IMF}}
\startdata
Model 1 & $1$--$100$ M$_\odot$ Salpeter & $0.1$--$100$ M$_\odot$ Salpeter \\
Model 2 & $35$--$100$ M$_\odot$ Salpeter & $0.1$--$100$ M$_\odot$ Salpeter \\
Model 3 & $100$--$260$ M$_\odot$ Salpeter & $0.1$--$100$ M$_\odot$ Salpeter \\
\enddata 
\end{deluxetable}

We always take the stellar IMF to have the Salpeter form, which is given by 
\begin{equation}
  \phi(m)=\phi_0\,m^{-2.3},
\end{equation}
where the constant $\phi_0$ is chosen such that 
\begin{equation}
  \int_{m_0}^{m_1}dm\,m\,\phi(m)=1\, \mathrm{M}_\odot. 
\end{equation}
We take $m_0=0.1$ M$_\odot$ and $m_1=100$ M$_\odot$ for
Population~I/II stars.  Population~III stars have different IMF in
each of our models, as we want to discuss the effect of Population~III
IMF on various quantities.  These are shown in Table
\ref{table:models}.  The stellar IMF in model 3 covers the pair
instability supernova range, whereas models 1 and 2 explore the effect
of AGB stars and core collapse supernovae respectively.  Each model is
calibrated separately to reproduce the cosmic star formation rate
(SFR) history, the IGM Thomson scattering optical depth to the last
scattering surface ($\tau_e=0.089\pm 0.014$
\citealt{2012arXiv1212.5226H}), and the hydrogen photoionization rate
in the IGM as measured from Ly$\alpha$ forest observations.

Our model requires stellar lifetimes because we do not assume
instantaneous recycling.  We take lifetimes for low and intermediate
mass stars ($0.1\,\mathrm{M}_\odot < M < 100\, \mathrm{M}_\odot$) from
\citet{1989A&A...210..155M}.  Lifetimes for stars with higher mass
(all of which are Population~III) are taken from
\citet{2002A&A...382...28S}.  The model also requires chemical and UV
photon yields of stars with different IMFs.
\citet{1995ApJS..101..181W} have calculated the chemical yields of
massive stars ($12\,\mathrm{M}_\odot < M < 40\, \mathrm{M}_\odot$)
with different metallicities ($Z=0$, $10^{-4}$, $0.01$, $0.1$, and $1$
times solar).  We use their results by interpolating between different
metallicities and extrapolating beyond the mass range for lower and
higher stellar masses.  Chemical yields of Population~III stars (which
are metal-free and high mass) are taken from the calculations of
\citet{2002ApJ...567..532H}.  Stellar spectra of Population~I/II stars
are calculated using {\scshape starburst99}
\citep{1999ApJS..123....3L, 2005ApJ...621..695V} with respective
metallicities.  Synthetic spectra of Population~III stars are taken
from \citet{2002A&A...382...28S}.

The transition from Population~III star formation to Population~II
star formation in any halo is implemented via a critical metallicity,
$Z_\mathrm{crit}$, with our fiducial value as $Z_\mathrm{crit}=10^{-4}$
\citep{2001MNRAS.328..969B, 2003Natur.425..812B, 2007MNRAS.380L..40F}.
When the ISM metallicity in a halo crosses $Z_\mathrm{crit}$, new stars
are formed according to a Population~II IMF.  We consider the effect
of changing the value of $Z_\mathrm{crit}$ below.

Note that in the chemical evolution model described above, we have
only accounted for core-collapse supernovae; the contribution of Type
Ia supernovae has been ignored.  As implicitly assumed in Equation
\ref{eqn:sfrz}, the core collapse supernova rate traces star formation
activity.  The Type Ia supernova rate has a more complex dependence on
the stellar IMF and star formation rate.

Classically, progenitors of SNe Ia are thought to be intermediate-mass
stars.  An unknown delay is expected between the death of the
progenitor star and the SN Ia explosion.  By using a similar model of
cosmic star formation as that presented here,
\citet{2006ApJ...647..773D} argued that the observed rate of SNe Ia
occurence suggests that the typical delay time is long ($\sim
3$--$3.5$ Gyr).  Using a different model, \citet{2004MNRAS.347..942G}
reached a similar conclusion, arguing that observations favour long
delay times ($\sim 1$ Gyr).  This is also in agreement with recent DLA
enrichment measurements at $z=4$--$5$ by \citet{2012ApJ...755...89R},
who report $\alpha$-enhancement of about 0.3 in these objects,
suggesting that these objects have not yet been enriched by SNe Ia.
(As we will discuss below in Section \ref{sec:discuss}, the effect of
dust depletion is expected to be low in these systems due to their low
metallicity.) If the SNe Ia delay times are indeed sufficiently long,
the results of this paper will not be affected, as we are interested
in DLA abundances at the highest redshifts ($z\gtrsim 4$).

However, \citet{2006MNRAS.370..773M} have argued that the observed
dependence of SNe Ia rate on the colours of parent galaxies, and
observed SNe Ia rates in radio-loud early-type galaxies favour a
bimodal delay time distribution, in which about half of SNe Ia explode
soon after the birth of their progenitor star (short delay time of
$\sim 100$ Myr), while the remaining half have long delay time of the
order of 3 Gyr.  In the context of DLAs, \citet{2003MNRAS.340...59C}
have also shown that when dust depletion is taken into account, most
of the claimed DLA $\alpha$-enhancements vanish, which suggests a role
of SNe Ia \citep[cf.][]{1996ApJ...462...57M}.  While dust is not
expected to play a major role in $z\sim 5$ DLAs, these prompt SNe Ia
with short delay times can have an effect on the results of our model.
Nevertheless, in this paper, we focus on the effect of Population~III
star formation on the abundance ratios in this paper.  We will
consider the effects of dust depletion and prompt SNe Ia in a future
work (Kulkarni et al. 2013, in preparation).  We conjecture here,
however, that while such prompt SNe Ia will affect our results, a
strong degeneracy with the effect of Population III IMF is unlikely as
the latter leave a sufficiently large signature in ratios between
$\alpha$-element abundances ($\sim 0.5$--$1$ dex at $z\sim 6$) .

\subsection{IGM metallicity}

Cosmological inflows bring material from the IGM into the ISM of
galaxies.  The extent to which this dilutes the metal content of the
ISM depends on the metallicity of the IGM.  We follow the model of
\citet{2006ApJ...647..773D} to calculate the average IGM metallicity
at any redshift.  This model divides cosmic baryons into three
reservoirs: (1) intergalactic medium (subscript ``IGM''), (2)
interstellar medium (``ISM''), and (3) stars (``$\mathrm{str}$'').  We
denote the mass densities of these as $M_\mathrm{IGM}$,
$M_\mathrm{ISM}$ and $M_\mathrm{str}$ respectively.  Note that
$M_\mathrm{ISM}$ and $M_\mathrm{str}$ are mass function weighted
integrals of $M_g$ and $M_*$ defined for each halo mass bin above.

The IGM mass density evolves according to
\begin{equation}
  \frac{dM_\mathrm{IGM}}{dt}=-a_b(t)+o(t),
\end{equation}
where $a_b$ is the \emph{total} rate of accretion of baryons from the
IGM on to halos of all masses, and $o(t)$ is the total rate of outflow
of baryons from halos into the IGM.  The stellar mass density evolves
as
\begin{equation}
  \frac{dM_\mathrm{str}}{dt}=\Psi(t)-e(t),
\end{equation}
where $\Psi$ is the total cosmic star formation rate, and $e$ is rate
of ejection of material of stars into the ISM via winds and
supernovae, which depends on the stellar IMF.  The total cosmic star
formation rate is given by 
\begin{equation}
  \Psi(t)=\int^{M_h^\mathrm{high}}_{M_h^\mathrm{low}} dM \dot M_*(M,t) N(M,t), 
\end{equation}
where $M_h^\mathrm{high}$ and $M_h^\mathrm{low}$ are the halo mass
limits considered in our calculation (as discussed above), $\dot
M_*(M,t)$ is star formation rate within a halo of mass $M$ at cosmic
time $t$ (given by Equation \ref{eqn:halosfr}), and $N(M,t)$ is the
halo mass function.  Lastly, the ISM mass density evolves as
\begin{equation}
  \frac{dM_\mathrm{ISM}}{dt} = -\frac{dM_\mathrm{IGM}}{dt} -
  \frac{dM_\mathrm{str}}{dt}.
\end{equation}

To calculate the IGM metallicity, in addition to the total masses, the
mass fraction of individual elements in each reservoir is also
evolved.  For an element $i$, the mass fraction in the IGM is
$X^\mathrm{IGM}_i=M^\mathrm{IGM}_i/M_\mathrm{IGM}$ and that in the ISM
is $X^\mathrm{ISM}_i=M^\mathrm{ISM}_i/M_\mathrm{ISM}$.  It can then be
shown that mass fractions in the IGM evolve as
\begin{equation}
  \frac{dX^\mathrm{IGM}_i}{dt}=o\left(X_i^\mathrm{ISM}-X_i^\mathrm{IGM}\right)/M_\mathrm{IGM},
\end{equation}
and those in the ISM evolve as
\begin{multline}
  \frac{dX^\mathrm{ISM}_i}{dt}=\left[a_b\left(X_i^\mathrm{IGM}-X_i^\mathrm{ISM}\right)\right.\\\left.+\left(e_i-eX_i^\mathrm{ISM}\right)\right]/M_\mathrm{ISM}.
\end{multline}
Here, $o$ is the outflow rate from the ISM and $e_i$ is the rate at
which metal $i$ is ejected by stars.  In order to completely specify
the model, all that is left to specify is (1) initial conditions for
above equations, and (2) method of calculation for the three
quantities $o$, $a_b$, and $e$.

The total baryon accretion rate is given by
\begin{eqnarray}
  a_b(t) &=& \Omega_b\left(\frac{3H_0^2}{8\pi G}\right)\left(\frac{dt}{dz}\right)^{-1}\left|\frac{df_{\mathrm{struct}}}{dz}\right| \nonumber \\ 
  &=& 1.2h^3M_\odot\mathrm{yr}^{-1}\mathrm{Mpc}^{-3}\left(\frac{\Omega_b}{0.044}\right)\nonumber\\
& &\times (1+z)\sqrt{\Omega_\Lambda+\Omega_m(1+z)^3}\left|\frac{df_{\mathrm{struct}}}{dz}\right|,
\end{eqnarray}
where $f_{\mathrm{struct}}$ is the fraction of total mass in star
forming halos.  It is given by
\begin{equation}
  f_{\mathrm{struct}}(z)=\frac{\int^\infty_{M_\mathrm{min}} dM M f_\mathrm{PS}(M,z)}{\int^\infty_0 dM M f_\mathrm{PS}(M,z)},
\end{equation}
and $f_\mathrm{PS}$ is the Press-Schechter mass function.

Outflows from the ISM into the IGM are
described using the following relation
\begin{equation}
  o(t)=\frac{2\epsilon}{v^2_\mathrm{esc}(z)}\int^{m_u}_{m_l} dm\,\phi(m)\cdot\Psi[t(z)-\tau(m)]\cdot E_\mathrm{kin}(m),
\end{equation}
where, as before, $\phi$ is the stellar IMF, and $E_\mathrm{kin}$ is
the kinetic energy released by the explosion of a star of mass $m$.
The quantity $v_\mathrm{esc}$ is the ``escape velocity,'' which is
calculated according to
  \begin{equation}
    v^2_\mathrm{esc}(z)=\frac{\int^\infty_{M_\mathrm{min}}dM
      f_\mathrm{PS}(M,z) M
      (2GM/R_\mathrm{vir})}{\int^\infty_{M_\mathrm{min}}dM
      f_\mathrm{PS}(M,z) M}.
  \end{equation}
The ejection of enriched gas from stars into the ISM is given by
\begin{equation}
  e(t)=\int_{m_l}^{m_u}dm\,\phi(m)\cdot\Psi[t(z)-\tau(m)]\cdot[m-m_r(m)],
\end{equation}
as above, where $m_\mathrm{u}$ is the upper limit on mass of stars
that explode and produce supernova remnants, and $m_r$ is the remnant
mass.

\subsection{Ionization and thermal evolution of the inhomogeneous IGM}

Our model for reionization and thermal history of the average IGM is
essentially that developed by \citet{2005MNRAS.361..577C}, with the
main difference being in implementation of sources.  In this model,
which matches a wide range of observational constraints, the IGM is
gradually reionized and reheated by star-forming galaxies between
$z\sim 20$ and $z\sim 6$.  We summarise the main elements of the model
here and refer the reader to that paper for details.

The model accounts for IGM inhomogeneities by adopting a lognormal
density distribution with the evolution of volume filling factor of
ionised hydrogen (H~\textsc{ii}) regions $Q_{\rm HII}(z)$ being
calculated according to the method outlined by
\citet{2000ApJ...530....1M}.  The volume filling factor of
H~\textsc{ii} regions evolves as
\begin{equation}
  \frac{d[Q_{\rm HII}F_M(\Delta)]}{dt} = \frac{\dot n_\nu}{n_H} - 
  Q_{\rm HII}\alpha_R(T)n_eR(\Delta)(1+z)^3,
\label{eqn:qhii}
\end{equation}
where $F_M(\Delta)$ is the mass fraction of the IGM occupied by
regions with density less than $\Delta$, $\dot n_\nu$ is the rate at
which ionizing photons are introduced in the IGM by galaxies, $n_e$ is
the electron number density, $\alpha_R(T)$ is the
temperature-dependent recombination rate and $R(\Delta)$ is the
clumping factor of the IGM, which is the related to the second moment
of the IGM density distribution.  The IGM mass fraction $F_M(\Delta)$
is given by
\begin{equation}
  F_M(\Delta_\mathrm{crit}) = \int_0^{\Delta_\mathrm{crit}} d\Delta \Delta P(\Delta),
\end{equation}
and the clumping factor is given by 
\begin{equation}
  R(\Delta_\mathrm{crit}) = \int_0^{\Delta_\mathrm{crit}} d\Delta \Delta^2 P(\Delta).
\end{equation}

Reionization is said to be complete once all low-density regions (say,
with overdensities $\Delta < \Delta_{\rm crit} \sim 60$) are ionised.
We refer to the redshift at which this happens as the redshift of
reionization, $z_\mathrm{reion}$.  We follow the ionization and
thermal histories of neutral and H~\textsc{ii} regions simultaneously
and self-consistently, treating the IGM as a multi-phase medium.  We
assume that helium is singly-ionized together with hydrogen and
contributes to the photoionization heating.  We do not consider the
double ionization of helium, which is thought to have occurred at
$z\sim 3$ \citep[e.g.,][]{2011ApJ...733L..24W}.

The rate of ionising photons injected by the galaxies into the IGM per
unit time per unit volume at redshift $z$ is denoted by $\dot
n_\mathrm{ph}(z)$, which is determined by the star formation rate.
Using our star formation model described above, we can write the rate
of emission of ionising photons per unit time per unit volume per unit
frequency range, $\dot n_\nu(z)$, as
\begin{equation}
\dot
n_\nu(z)=f_\mathrm{esc}\left[N^\mathrm{II}_\gamma(\nu)\Psi^\mathrm{II}(z)+N^\mathrm{III}_\gamma(\nu)\Psi^\mathrm{III}(z)\right]
\label{nnu}
\end{equation}
where $f_\mathrm{esc}$ is the escape fraction of UV photons, and
$N^\mathrm{II}_\gamma(\nu)$ and $N^\mathrm{III}_\gamma(\nu)$ are the
total number of ionising photons emitted per unit frequency range per
unit stellar mass from Population~I/II and Population~III star forming
halos respectively.  These quantities are calculated by integrating
the stellar spectra over an appropriate range.  $\Psi^\mathrm{II}(z)$
and $\Psi^\mathrm{III}(z)$ are the cosmic Population~I/II and
Population~III star formation rates.  At any redshift,
\begin{equation}
\Psi(z) = \Psi^\mathrm{II}(z) + \Psi^\mathrm{III}(z).
\label{sfr:tot}
\end{equation}
The total rate of emission of ionising photons per unit time per unit
volume, $\dot n_\mathrm{ph}(z)$, is obtained by simply integrating 
Equation (\ref{nnu}) over the correct frequency range.  The hydrogen
photoionization rate is then given by 
\begin{equation}
  \Gamma_\mathrm{HI}(z)=(1+z)^3\int_{\nu_0}^\infty d\nu \lambda(z,\nu) \dot n_\nu(z) \sigma(\nu),
\end{equation}
where $\sigma(\nu)$ is the hydrogen photoionization cross-section, and
$\lambda(z,\nu)$ is the redshift-dependent mean free path of UV
photons.  The mean free path is governed by the distribution of high
density neutral clumps and reproduces the observed density of
Lyman-limit systems at low redshift.  The photoionization rate is used
with the average radiative transfer equation to calculate the
evolution of $Q_{\rm HII}(z)$ in the pre-reionization universe, and
that of $\Delta_\mathrm{crit}(z)$ in the post-reionization universe.
Note that this approach assumes ionization equilibrium and local
absorption of UV photons. 

The mean free path is defined as the average distance between high
density regions of the inhomogeneous IGM.  It can be written as
\citep{2005MNRAS.361..577C}
\begin{equation}
  \lambda(z) = \frac{\lambda_0}{[1-F_V(\Delta_\mathrm{crit})]^{2/3}},
\end{equation}
where $F_V(\Delta_\mathrm{crit})$ is the volume fraction of the IGM
occupied by regions with density less than $\Delta_\mathrm{crit}$,
given by
\begin{equation}
  F_V(\Delta_\mathrm{crit}) = \int_0^{\Delta_\mathrm{crit}} d\Delta P(\Delta),
\end{equation}
where $P(\Delta)$ is the probability distribution function of IGM
overdensities, taken to have a lognormal form
\citep{2000ApJ...530....1M, 2005MNRAS.361..577C}.  We take $\lambda_0$
to be proportional to the Jeans length (Equation \ref{eqn:jeans}),
\begin{equation}
  \lambda_0=\lambda_{\mathrm{mfp,0}}r_J,
\end{equation}
and constrain the constant of proportionality,
$\lambda_{\mathrm{mfp,0}}$, using the number of Lyman-limit systems at
low redshift, which is related to the mean free path by
\begin{equation}
  \frac{dN_\mathrm{LLS}}{dz}=\frac{c}{\sqrt(\pi)\lambda(z)H(z)(1+z)}.
\end{equation}
We assume that the mean free path is independent of the photon energy
up to the ionization threshold of He~\textsc{ii} and is proportional
to $(\nu/\nu_\mathrm{HeII})^{1.5}$ at higher energies
\citep{2005MNRAS.361..577C}.

The IGM temperature evolves as
\begin{multline}
  \frac{dT}{dt}=-2H(z)T-\frac{T}{1+X_\mathrm{HI}}\frac{dX_\mathrm{HI}}{dt}\\+\frac{2}{3k_Bn_b(1+z)^3}\frac{dE}{dt},
\end{multline}
where $dE/dt$ is the total photoheating rate and $X_\mathrm{HI}$ is
the H~\textsc{i} fraction.  The photoheating term accounts for heating
by the UV background and cooling due to recombinations and Compton
scattering of CMB photons.  The IGM temperature governs the thermal
feedback on galaxy formation by stopping gas inflow on halos with mass
less than $M_\mathrm{min}(z)$.  We take this to be the Jeans mass,
given by
\begin{equation}
  M_\mathrm{min}=\frac{4}{3}\pi r_J^3,
\end{equation}
where $r_J$ is the comoving Jeans length
\begin{equation}
  r_J = \left[\frac{2k_BT\gamma}{8\pi G \rho_c m_p \Omega_m (1+z)}\right]^{1/2},
  \label{eqn:jeans}
\end{equation}
where $\gamma$ is the adiabatic index and $m_p$ is the mass of the
proton.  Note that before reionization is complete, the H~\textsc{ii}
regions are distinct.  During this epoch, the temperature in
H~\textsc{ii} regions is different from that in H~\textsc{i} regions.
The average temperature is then given by the mean of these two
temperatures, weighted by $Q_{\rm HII}(z)$.  Similarly, both regions
have different $M_\mathrm{min}$.  This is relevant in understanding
the role that $M_\mathrm{min}$ plays in our star formation
prescription above.

Note that in all of the above, we evolve two distinct sets of haloes
corresponding to haloes forming in H~\textsc{i} and H~\textsc{ii}
regions respectively.  Each set of haloes is affected by
photoionization feedback differently as the temperatures of
H~\textsc{i} and H~\textsc{ii} regions are different.  Any average
property, such as the cosmic SFR, at a given redshift is an average of
these two sets weighted by the volume filling fraction of H~\textsc{i}
and H~\textsc{ii} regions.  This distinction is unimportant at
redshift $z<z_\mathrm{reion}$, when there are no H~\textsc{i} regions
anymore.

Given the above model, we obtain best-fit parameters by comparing with
various observations such as the redshift evolution of photoionization
rate obtained from the Ly$\alpha$ forest \citep{2007MNRAS.382..325B},
the electron scattering optical depth \citep{2012arXiv1212.5226H}, and
the total baryon fraction in structures at $z=0$
\citep{2004ApJ...616..643F}.  These parameters are summarised in Table
\ref{table:params} along with their fiducial values and the
constraints used to obtain them.

\begin{deluxetable}{ccc}
\tablecaption{Free parameters in our model\label{table:params}}
\tablehead{\colhead{Parameter} & \colhead{Fiducial value} & \colhead{Constraint}}
\startdata
$f_\mathrm{esc}$ & 0.04 & $\tau_e$ and $\Gamma_\mathrm{HI}$\\ 
$f_*$ (Pop.~II) & 0.02 & $\tau_e$ and $\Gamma_\mathrm{HI}$\\ 
$f_*$ (Pop.~III) & 0.04 & $\tau_e$ and $\Gamma_\mathrm{HI}$\\ 
$\lambda_\mathrm{mfp,0}$ & $1.7\times 10^{-3}$ & $dN_\mathrm{LLS}/dz$\\ 
$\epsilon$ & $1.0\times 10^{-6}$ & $f_{b,\mathrm{struct}}(z=0)$\\
$f_\mathrm{gas,in}$ & 0.7 & \citet{2010ApJ...718.1001B}\\ 
$\zeta$ & Eq. (\ref{eqn:zeta}) & \citet{2012ApJ...753...16K}\\ 
$Z_\mathrm{crit}$ & $10^{-4}$ $Z_\odot$ & $-$\\ 
\enddata 
\end{deluxetable}

Before moving on to the results of our model, we compare our galaxy
formation model with conventional semi-analytical models of galaxy
formation used in the literature \citep[e.g.,][]{1999MNRAS.303..188K,
  2000MNRAS.319..168C, 2004MNRAS.349.1101D, 2006RPPh...69.3101B}.  The
basic approach in these models is two-step: first, halo merger trees
are constructed from N-body cosmological simulations; next, baryonic
physics is calculated in each individual halo using prescriptions
similar to those described above.  When haloes merge, the stellar,
gas, and metal masses are suitably combined, possibly with some extra
prescriptions for events such as starbursts or active galactic nuclei
(AGN).  Our approach in modeling galaxy formation in this paper is
also two-step.  However, instead of drawing halo merger trees from
N-body simulations, we only draw the average mass assembly histories
of haloes.  This is a significant reduction in information, as the
average mass at redshift $z$ of a halo with mass $M$ at $z=0$ is the
mean of the masses all progenitors at redshift $z$ of all such haloes
in the simulation box (see \citealt{2010MNRAS.406.2267F} for details
of how this is done).  In particular, by considering only the mean
assembly histories, we lose any information regarding the intrinsic
scatter in any property (such as halo mass, stellar mass, gas mass, or
chemical abundance), even for fixed halo mass.  Nonetheless, this
simplification affords us the ability to calculate the IGM thermal and
ionization histories and the evolution of H~\textsc{i} and
H~\textsc{ii} regions in the IGM by solving the average radiative
transfer equation (Equation \ref{eqn:qhii}).  It is not possible to do
this in conventional merger-tree-based semi-analytic models.

Although our approach is different from conventional semi-analytic
models, we expect our results to correspond to averaged results of
those models.  Thus, e.g., the stellar mass at a given redshift of a
given halo of mass, say, $M$ in our model, corresponds to the mean
stellar mass of all haloes at that redshift in a merger-tree-based
model.  This can be seen from the agreement between our model
predictions and observational data for average quantities such as
cosmic star formation rate, gas phase metallicity, and
mass-metallicity relation, as we discuss below.

\section{Results}
\label{sec:results}

\subsection{Global properties of the chemical evolution models}

\begin{figure}
  \begin{center}
    \includegraphics[scale=0.5]{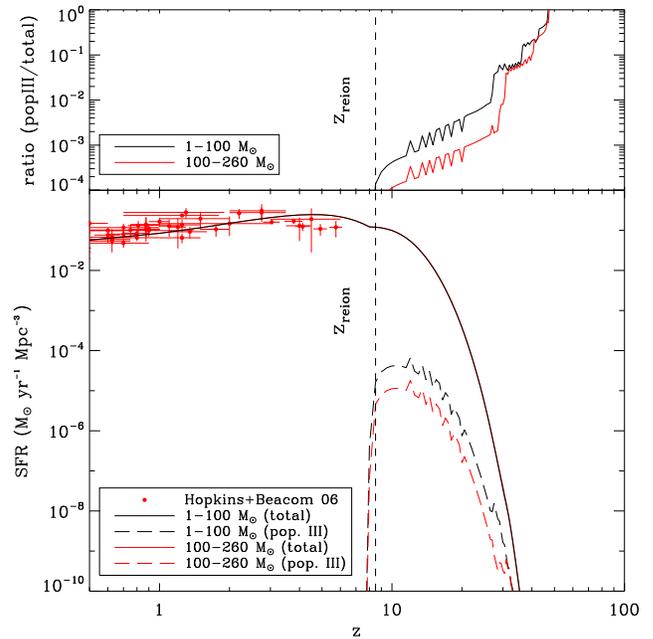}
  \end{center}
  \caption{Evolution of the cosmic star formation rate (SFR) density
    in our models.  In the bottom panel, solid and dashed curves show
    model prediction for the total and Population~III SFR densities
    respectively.  Black curves correspond to 1--100 M$_\odot$
    Population~III IMF (model 1) and red curves to 100--260 M$_\odot$
    Population~III IMF (model 3).  Black and red solid curves are
    found to overlap at all redshifts.  Data points are observational
    measurements from a compilation by \citet{2006ApJ...651..142H}.
    In the top panel, we show the evolution of the ratio of
    Population~III SFR to the total SFR.  fractional contribution of
    Pop.~III SFR to the total SFR.  Dashed vertical line shows the
    redshift of reionization.  The oscillatory features in the curves
    are numerical effects.}
  \label{plot:sfr}
\end{figure}

Our model for galaxy and IGM evolution is constrained to produce the
observed star formation rate (SFR) density evolution, the fraction of
total baryon density in haloes at low redshift, and the reionization
history of the IGM as measured by the Thomson scattering optical depth
to the last scattering surface ($\tau_e$) and the hydrogen
photoionization rate ($\Gamma_\mathrm{HI}$).  Figure \ref{plot:sfr}
shows the evolution of the cosmic SFR density.  The solid curves show
the prediction of our model and the red data points show observational
measurements from a compilation by \citet{2006ApJ...651..142H}.  All
three models reproduce the observed rise in the SFR between $z=0$ and
$z\sim 3$.  The total SFR prediction of the three models are virtually
identical.  As we discuss in detail below, this is because the
contribution from Population~III stars is very small.  However, given
the error bars on the measurements and the simplicity of our model, we
consider the match to be adequate.  Note that recent measurements of
the SFR density at $z\sim 7$--$10$ by \citet{2010ApJ...709L.133B} and
\citet{2013ApJ...763L...7E} are smaller than our model prediction at
these redshifts.  These measurements report values of $\lesssim
10^{-2}$ M$_\odot$ yr$^{-1}$ Mpc$^{-3}$ at $z\sim 8$ and $\lesssim
10^{-3}$ M$_\odot$ yr$^{-1}$ Mpc$^{-3}$ at $z\sim 10$.  Our values at
both these redshifts $\gtrsim 10^{-2}$ M$_\odot$ yr$^{-1}$ Mpc$^{-3}$.
However, this discrepancy is because the SFR estimates of
\citet{2010ApJ...709L.133B} and \citet{2013ApJ...763L...7E} do not
count the large contribution to the SFR by faint galaxies (which are
more numerous).  These estimates are obtained by integrating the UV
luminosity functions down to $0.05L^*_{z=3}$.  This limit corresponds
to an AB magnitude of about $-18$ at $1500$\AA, while we find galaxies
down to $\gtrsim -17$ in our model.  In a forthcoming paper (Kulkarni
et al., in prep.), we analyse this issue in detail and show that our
model predictions are in good agreement with high-redshift SFR
estimates when the faintness limit of \citet{2010ApJ...709L.133B} and
\citet{2013ApJ...763L...7E} is applied.

\begin{figure*}
  \begin{center}
  \begin{tabular}{rl}
    \includegraphics[scale=0.5]{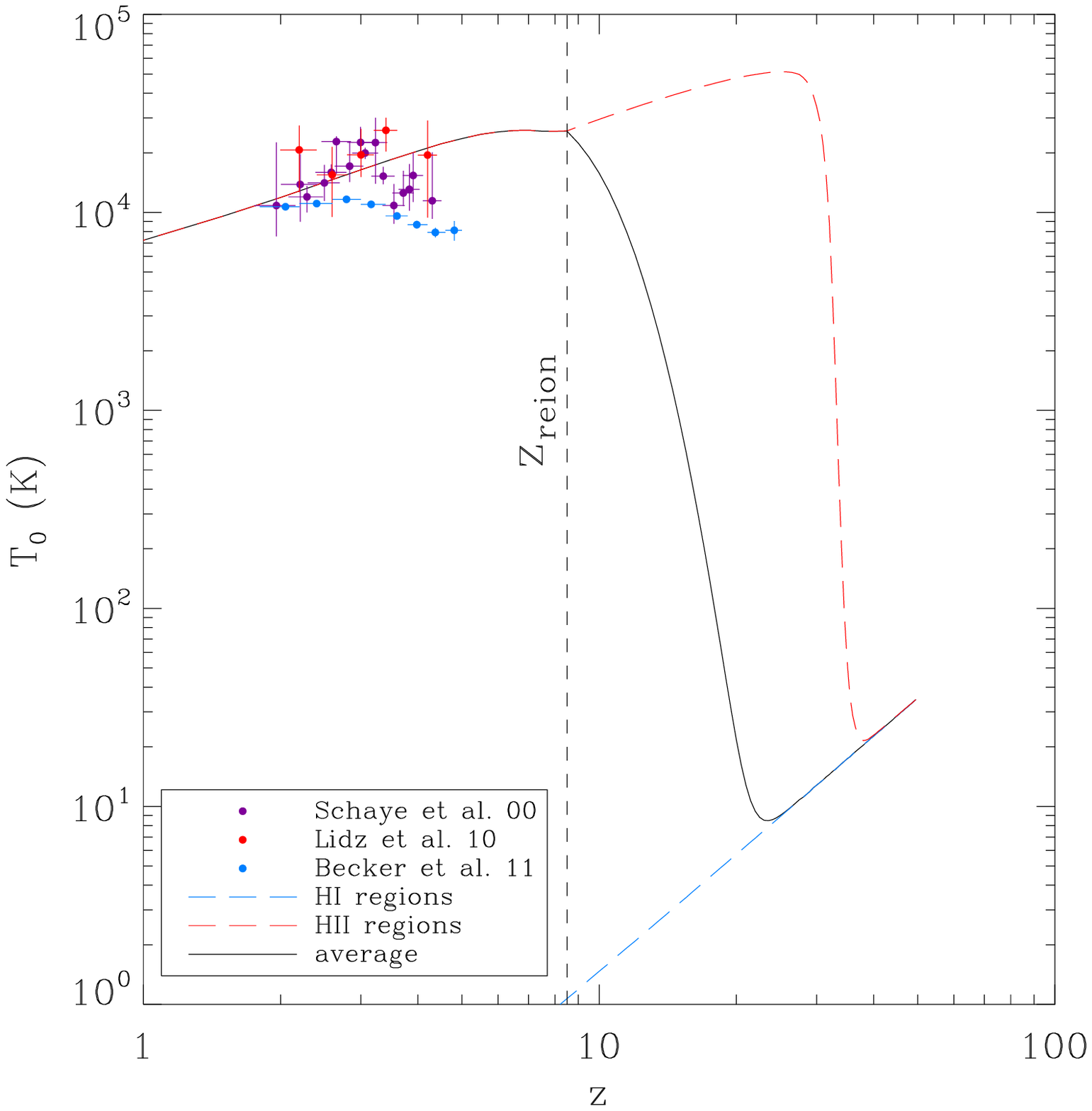} &
    \includegraphics[scale=0.5]{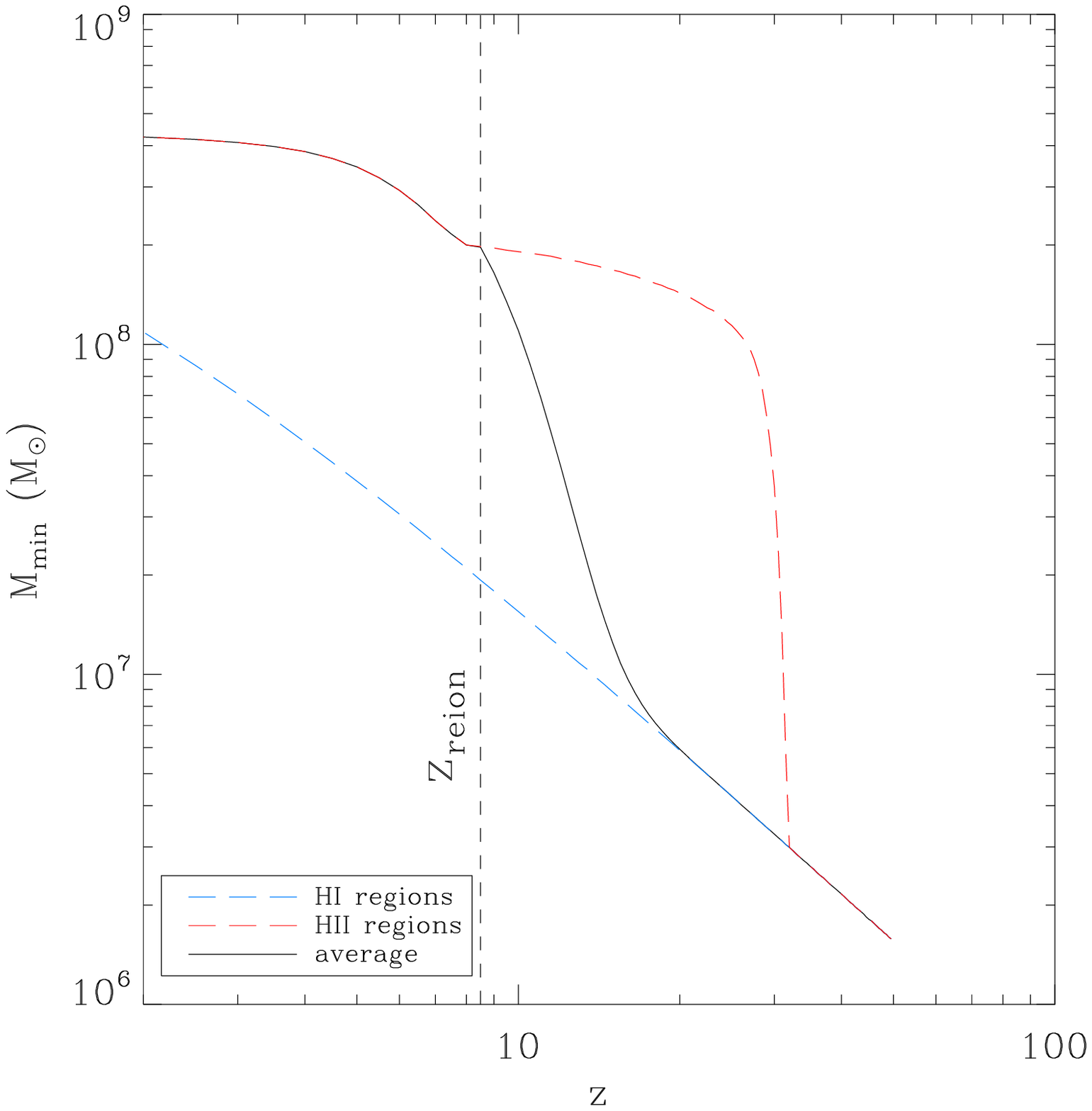} \\
  \end{tabular}
  \end{center}
  \caption{Evolution of IGM temperature (left panel) and the filtering
    mass (right panel) in our models, shown here for model 3. In both
    panels, the blue dashed curve shows the evolution in H~\textsc{i}
    regions and the red dashed curve shows the evolution in
    H~\textsc{ii} regions.  The average is shown by the solid black
    curve in both panels.  Dashed vertical lines in both panels shows
    the redshift of reionization.  Data points in the left panel are
    observational measurements of the IGM temperature from the
    Lyman-$\alpha$ forest \citep{2000MNRAS.318..817S,
      2010ApJ...718..199L, 2011MNRAS.410.1096B}. }
  \label{plot:mmin}
\end{figure*}

Figure \ref{plot:mmin} shows the thermal evolution of the IGM in our
models in comparison with some observational measurements
\citep{2000MNRAS.318..817S, 2010ApJ...718..199L, 2011MNRAS.410.1096B}.
The solid curve in left panel shows the evolution of the temperature
of the IGM at mean density ($T_0$), which is the temperature usually
measured from absorption spectra of high redshift quasars.  At
redshift $z\gtrsim 20$, $T_0$ decreases adiabatically $\propto
(1+z)^2$. However, as the H~\textsc{ii} regions around first galaxies
form, this temperature gradually rises to about $2\times 10^4$ K at
$z=z_\mathrm{reion}\sim 6$.  Once reionization is complete, only the
highest density regions in the IGM are left to be ionized.  Thus the
mean density regions evolve unaffected and $T_0$ then decreases to
$\sim 10^4$ K.  At redshift higher than $z_\mathrm{reion}$, $T_0$ is
the mean of temperatures in H~\textsc{i} and H~\textsc{ii} regions,
weighted by the volume filling factor.  As a result, at these
redshifts temperatures in these regions are more relevant.  In Figure
\ref{plot:mmin} (left panel), the red dashed curve shows the
temperature evolution in H~\textsc{ii} regions, and the blue dashed
curve shows the same in H~\textsc{i} regions.  The average
H~\textsc{ii} region temperature climbs to a high value of a few times
$10^4$ K at $z\sim 20$ when the first sources turn on.  However, this
does not affect $T_0$ at these redshifts as the H~\textsc{ii} region
volume filling factor is small.  The average temperature in
H~\textsc{i} regions continues to decrease with time due to adiabatic
cooling.  Note that as we do not model He~\textsc{ii} reionization in
this paper, our ability in matching the observational measurements is
restricted.  All observational measurements are made in the redshift
range $z=2$--$4$, which is believed to be the epoch of He~\textsc{ii}
reionization.  We also note in passing that, while the only source of
heat in our model is photoionizations, the temperature measurements in
the redshift range $z=2$--$4$ are not precise enough to rule out
other, exotic, sources \citep{2004MNRAS.348L..43B,
  2009ApJ...694..842M, 2012MNRAS.422.3019C, 2012MNRAS.423....7M,
  2012ApJ...757L..30R, 2012MNRAS.424.1723G}.

Recall that here the volume filling factor of H~\textsc{ii} regions is
calculated according to Equation (\ref{eqn:qhii}).  Present
observational constraints on this quantity are not very strong.
However, several upper limits have been suggested in the literature.
These include: (1) a lower limit of $Q_\mathrm{HII}\gtrsim 0.6$
derived by \citet{2010ApJ...723..869O} at $z=6.6$ from the evolution
of the number density of Lyman-$\alpha$ emitting galaxies, (2) an
upper limit on $Q_\mathrm{HI}$ at $z=5.5$ and $6$ measured by
\citet{2010MNRAS.407.1328M} by counting dark pixels in quasar spectra,
(3) a lower limit of $Q_\mathrm{HII}\gtrsim 0.5$ inferred by
\citet{2007MNRAS.381...75M} at $z=6.6$ from the lack of an increase in
the clustering of Lyman-$\alpha$ emitting galaxies, (4) an upper limit
of $Q_\mathrm{HII}\lesssim 0.1$ inferred from the proximity zone of a
$z=7.1$ quasar by \citet{2011MNRAS.416L..70B}, and (5) an upper limit
of $Q_\mathrm{HI}\lesssim 0.5$ inferred by \citet{2008MNRAS.388.1101M}
from the red damping wing of the Ly$\alpha$ absorption line in the
spectrum of a $z=6.3$ gamma-ray burst.  Our model is consistent with
all of these constraints as in our model reionization is complete by
redshift $z=8$.

The thermal evolution of the IGM dictates the evolution of
$M_\mathrm{min}$, the minimum mass of star-forming haloes, which is
shown in the right panel in Figure \ref{plot:mmin}.  This quantity
implements photoionization feedback in our model.  As described in the
previous section, we calculate $M_\mathrm{min}$ using a Jeans
criterion.  At a given redshift the minimum mass of star-forming
haloes is either given by the atomic cooling threshold of $10^4$ K, or
by the local Jeans mass, whichever is higher.  Other models of
$M_\mathrm{min}$ are present in the literature, which take into
account the full thermal history of the IGM instead of the
instantaneous temperature \citep{2000ApJ...542..535G,
  2008MNRAS.390..920O}.  We ignore these improvements and focus on the
Jeans method for simplicity; this approximation does not affect our
results strongly.  The solid curve in Figure \ref{plot:mmin} (left
panel) shows the average $M_\mathrm{min}$ evolution.  It increases
monotonically with decreasing redshift and is $\sim 2\times 10^8$
M$_\odot$ at $z=z_\mathrm{reion}$.  It is $\sim 6\times 10^{8}$
M$_\odot$ at redshift 1.  It is important to note that this is only
the average $M_\mathrm{min}$; the minimum mass is different in
H~\textsc{i} and H~\textsc{ii} regions.  These are shown by the blue
and red curves in Figure \ref{plot:mmin}.  Thus, galaxies forming in
these regions are affected differently.  The minimum mass in
H~\textsc{ii} region grows to about $10^8$ M$_\odot$ as soon as the
first sources are turned on and the temperature of these regions is
boosted to $\sim 10^4$ K.  On the other hand, the minimum mass in
H~\textsc{i} regions increases much more slowly with decreasing
redshift as it is always determined by the atomic cooling threshold
alone.  The relative contribution of the two regions to the solid
black curve in Figure \ref{plot:mmin} is determined by the filling
factor of H~\textsc{ii} regions.  Of course, this distinction is not
important in the post-reionization phase, where there are no
H~\textsc{i} regions.

\begin{figure}
  \begin{center}
    \includegraphics[scale=0.5]{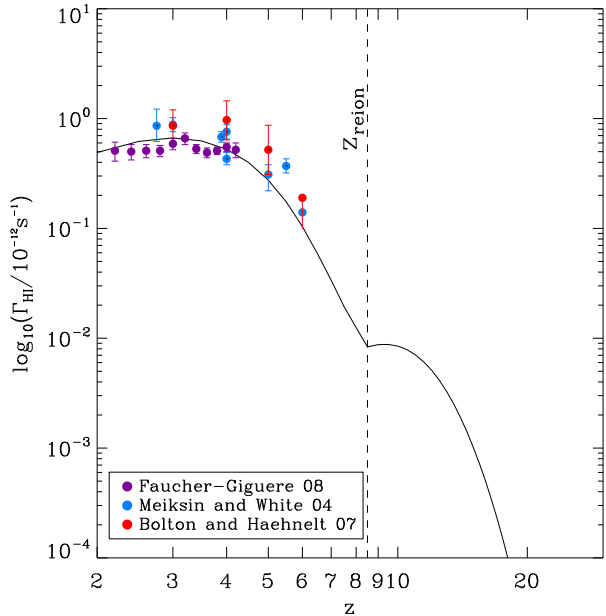}
  \end{center}
  \caption{Hydrogen photoionization rate in our models (solid curve),
    compared with measurements from the Ly$\alpha$ forest.  Blue data
    points show measurements by \citet{2004MNRAS.350.1107M}, red
    points are by \citet{2007MNRAS.382..325B}, and magenta points are
    by \citet{2008ApJ...688...85F}.  Dashed vertical line shows the
    redshift of reionization.}
    \label{plot:gpi}
\end{figure}

As seen in the thermal evolution in Figure \ref{plot:mmin}, IGM
reionization in our model is gradual.  This gradual change in the
ionization state of the IGM helps us simultaneously reproduce the two
most robust observational constraints on reionization: (1) the
observed Thomson scattering optical depth to the last scattering
surface $\tau_e=0.089\pm 0.014$ \citep{2012arXiv1212.5226H}, and (2)
the measurements of the hydrogen photoionization rate,
$\Gamma_\mathrm{HI}$, from observations of the Ly$\alpha$ forest
\citep{2004MNRAS.350.1107M, 2007MNRAS.382..325B, 2008ApJ...688...85F}.
The hydrogen photoionization rate in our model 3 is shown in Figure
\ref{plot:gpi}; our other models (1 and 2) give very similar results.
The model prediction matches very well with the observational
measurements.  The photoionization rate increases rapidly as UV photon
sources build up at $z>z_\mathrm{reion}$.  There is a sudden jump at
$z=z_\mathrm{reion}$ when different H~\textsc{ii} regions overlap.
This is because at this redshift H~\textsc{ii} regions overlap and a
given point in the IGM starts ``seeing'' multiple sources, which
rapidly enhances the UV photon mean free path, thereby affecting the
photoionization rate.

\begin{figure}
  \begin{center}
    \includegraphics[scale=0.5]{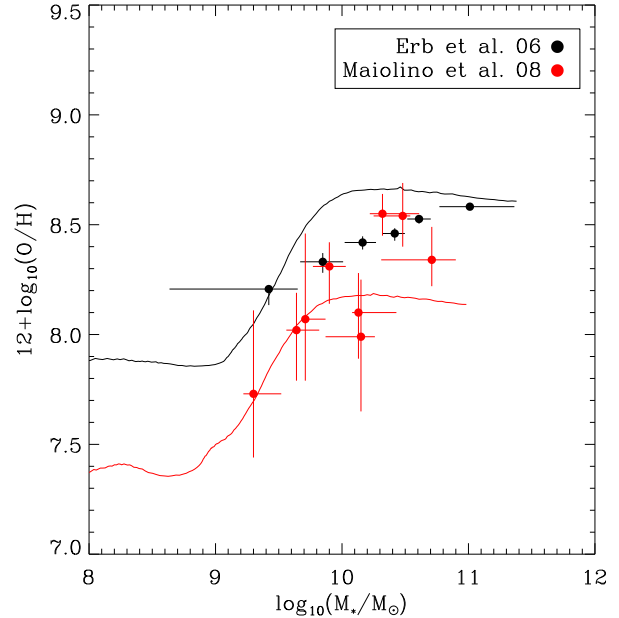}
  \end{center}
  \caption{The mass-metallicity relation prediction from our models
    (solid curves) compared with observational measurements (data
    points).  Measurements by \citet{2006ApJ...644..813E} at $z=2.27$
    are shown in black and those by \citet{2008A&A...488..463M} at
    $z=3.7$ are shown in red.  Note that the amplitudes of
    mass-metallicity relations are uncertain and have been rescaled as
    described in the text.}
    \label{plot:mzr}
\end{figure}

We now turn to the mass-metallicity relation prediction in our models,
which is shown in Figure \ref{plot:mzr}.  The mass-metallicity
relation can be easily predicted in our model because for each halo
mass, we calculate the stellar mass and gas-phase abundances of
various metals at various redshifts, as explained above in section
\ref{sec:model}.  In particular, in Figure \ref{plot:mzr}, we compare
the stellar mass with the oxygen abundance.  We find that the model
predictions match quite well with observations at $z=2.27$
\citep{2006ApJ...644..813E} and $z=3.7$ \citep{2008A&A...488..463M}.
The oxygen abundance increases with stellar mass for galaxies in the
intermediate mass range.  At the high mass end ($M_*>10^{11}
\mathrm{M}_\odot$) we find a slow decline in the oxygen abundance.
This is because the gas accretion rate in these haloes is high, which
results in dilution of their ISMs.  Also, it is difficult for outflows
to remove gas from the deeper potential wells of these halos.  Note
that the measurements of the mass-metallicity relationship are quite
uncertain \citep{2008ApJ...681.1183K, 2012ApJ...753...16K}.
Therefore, to compare our estimate of metallicity to the observations
we force the zero point of the predictions to match with that of the
observations at $z=2.27$.  The model mass-metallicity relationship has
a slightly different slope than the observational points at $z=2.27$.
However, given the simplicity of our model, and our focus on ratios of
abundances in this work, we take the level of agreement to be
sufficient.  We will also see below that when the average ISM
metallicity of galaxies is averaged agrees perfectly with the mean ISM
metallicity inferred from observations of damped Ly$\alpha$ absorbers
(DLAs).

In summary, our average galaxy formation and chemical evolution model
is consistent with several observations, including the observed
Thomson scattering optical depth to the last scattering surface, the
observed hydrogen photoionization rate inferred from the Ly$\alpha$
forest, the slope of the mass-metallicity relation of galaxies at
various redshifts, and the evolution of cosmic star formation rate
density.  It also agrees well with the low-mass slope, characteristic
mass, and normalization of the stellar-to-halo-mass relation, and, as
we will show below, the evolution of the mean ISM metallicity inferred
from observations of DLAs.

\subsection{Effect of the Population~III stellar IMF}
\label{sec:pop3effect}

\begin{figure*}
  \begin{center}
    \includegraphics{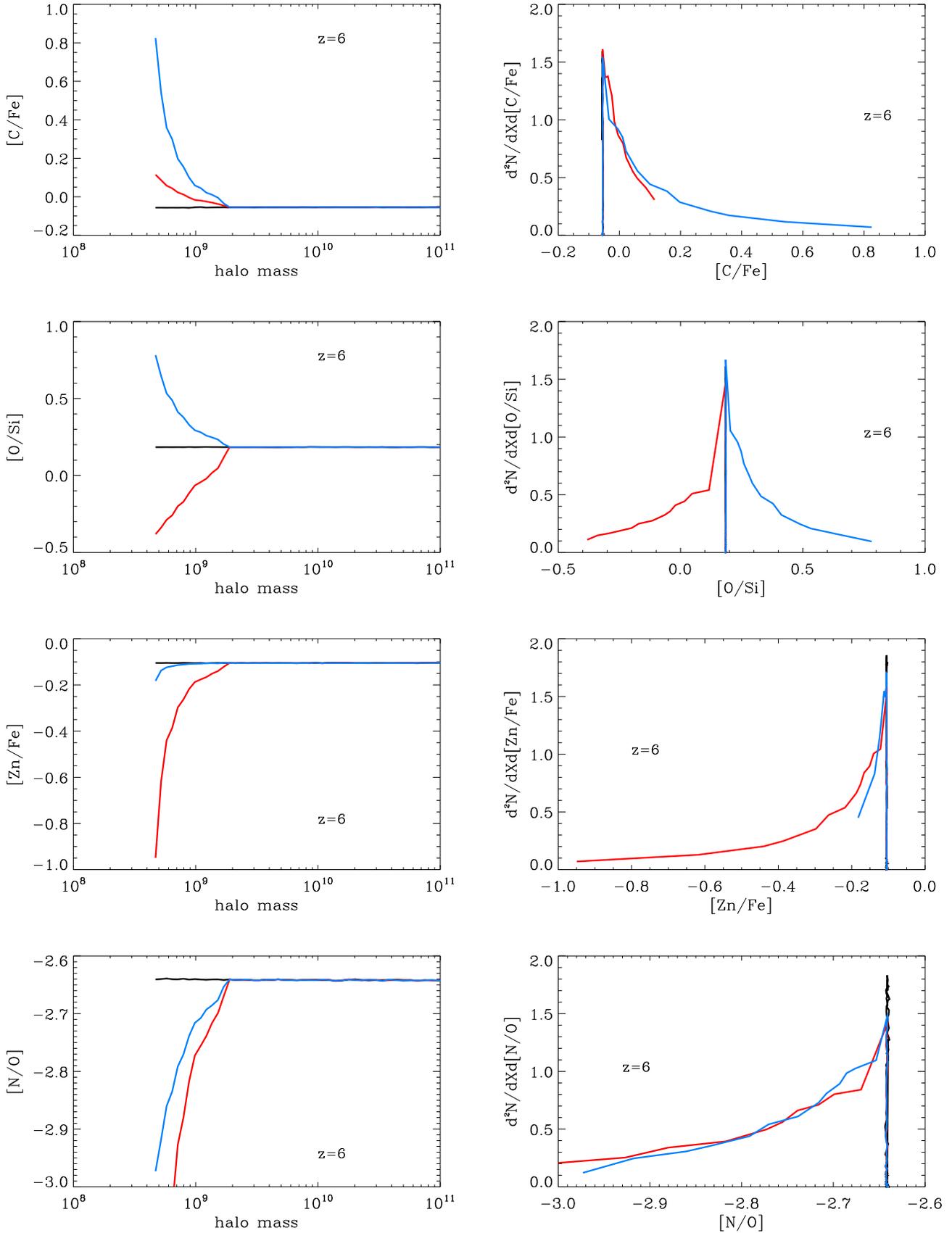}
    \caption{Abundance ratios and DLA distribution functions predicted
      by our model at $z=6$.  Panels in the left column shows the
      dependence of abundance ratios of halo mass in our model, while
      those in the right column show the predicted distribution of DLA
      abundance ratios.  Black, blue and red curves correspond to
      models 1, 2, and 3 respectively.}
    \label{plot:abr_z6}
  \end{center}
\end{figure*}

\begin{figure*}
  \begin{center}
    \includegraphics{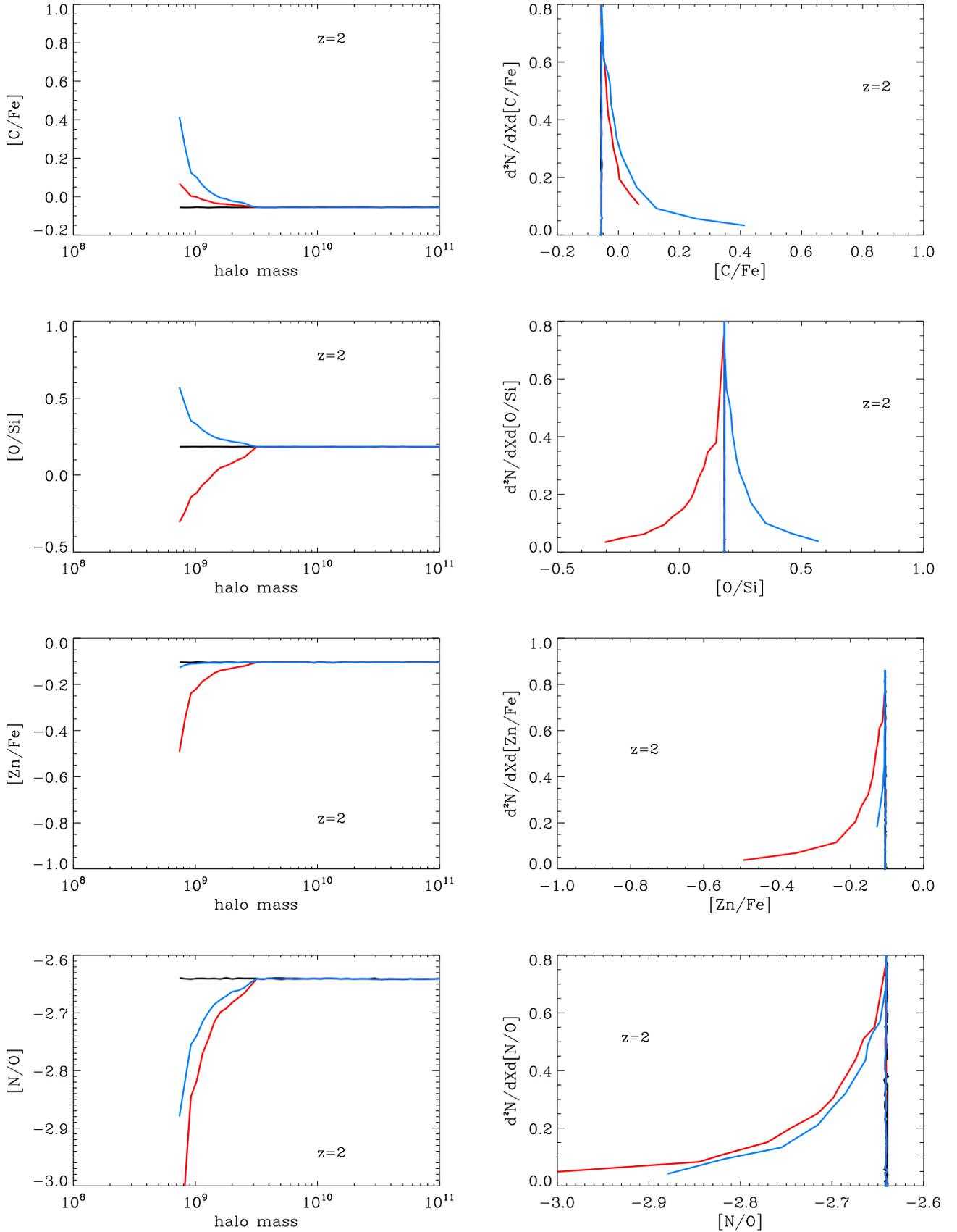}
    \caption{Abundance ratios and DLA distribution functions predicted
      by our model at $z=2$, shown on the same scale as Figure
      \ref{plot:abr_z6}.  Columns are same as Figure
      \ref{plot:abr_z6}.  Black, blue and red curves correspond to
      models 1, 2, and 3 respectively.  We see that the spread in the
      values of abundance ratios is reduced considerably at $z=2$ as
      compared to $z=6$.}
    \label{plot:abr_z2}
  \end{center}
\end{figure*}

\begin{deluxetable}{lccc}
\tablecaption{Logarithmic yields of different species in our three
  models, obtained by using Equation (\ref{eqn:yields}).  Note that
  some yields, e.g., Fe and Si, do not have a monotonic dependence on
  stellar mass.\label{table:yields}} \tablehead{\colhead{Species} &
  \colhead{Model 1} & \colhead{Model 2} & \colhead{Model 3}}
\startdata C & $-2.34$ & $-1.72$ & $-1.51$ \\ N & $-5.42$ & $-6.12$ &
$-6.49$ \\ O & $-1.59$ & $-0.94$ & $-0.58$ \\ Si & $-3.04$ & $-3.17$ &
$-1.29$ \\ Fe & $-2.71$ & $-4.31$ & $-1.87$ \\ Zn & $-5.95$ & $-7.16$
& $-6.98$ \\ \enddata
\end{deluxetable}

To study how the Population~III stellar IMF affects above results, we
now consider three different Population~III stellar IMFs shown in
Table \ref{table:models}.  Several early studies predicted that
Population~III stars have a characteristic mass of a few hundred
M$_\odot$ \citep[e.g.,][]{2002Sci...295...93A}.  In recent years, this
prediction has come down in the range of 20--100 M$_\odot$
\citep[e.g.,][]{2009Sci...325..601T}.  Recently,
\citet{2012arXiv1203.6842D} argued that in the absence of metal-line
cooling, dynamical effects can still lead to fragmentation in
proto-stellar gas clouds.  In their simulations, this resulted in
Population~III stars with masses as low as 0.1 M$_\odot$.  (This
picture predicts the existence of many Population~III stars surviving
in the Galaxy till the present day.)  Enrichment beyond
$Z_\mathrm{crit}$ merely results in a change in the slope of the IMF,
not in its mass values.  Given this lack of certainty in our knowledge
of the Population~III IMF, in this work we choose to work with three
different IMFs: 1--100 M$_\odot$ Salpeter, 35--100 M$_\odot$ Salpeter,
and 100--260 M$_\odot$ Salpeter.  In each case, our Population~II IMF
is fixed to 0.1--100 M$_\odot$ Salpeter.  Each of these three models
is calibrated independently, as described above.  Our goal is to
understand how the difference in these Population~III scenarios is
reflected in chemical evolution of galaxies.  The 1--100 M$_\odot$ and
100--260 M$_\odot$ IMFs are selected to represent two extreme
possibilities: if metal-free gas can fragment due to dynamical effects
the IMF will favour small stellar masses, but in the absence of
fragmentation Population~III stars could have masses of the order of
$\sim 100$ M$_\odot$.  The intermediate (35-100 M$_\odot$) IMF is
chosen to consider the effect of core collapse supernovae.  If the
distinct chemical signatures of these three IMFs turn out to be
observable in high redshift galaxies, then such observations could act
as a probe of the Population~III IMF.

The contribution of Population III stars to the total cosmic SFR
density is shown in Figure \ref{plot:sfr}, in which the bottom panel
shows the evolution of the Population~III SFR ($\Psi^\mathrm{III}(z)$)
and the top panel shows the fractional contribution to the total SFR
($\Psi^\mathrm{III}(z)/\Psi(z)$).  Evidently, Population~III stars
contribute very little in terms of SFR.  Furthermore, any
Population~III contribution is essentially zero by redshift $z\sim 7$.
The initial burst of Population~III stars is sufficient to enhance the
ISM metallicity beyond $Z_\mathrm{crit}$ in any halo mass bin.  When
this happens, the corresponding galaxy stops forming Population~III
stars.  Since most of the mass in the universe is contained in $M_*$
haloes, the globally averaged Population~III SFR starts declining as
soon as these haloes cross the $Z_\mathrm{crit}$ threshold.  Since the
IMF in model 1 has a lower metal yield than that in model 3, haloes
take longer to cross the $Z_\mathrm{crit}$ threshold.  Therefore, the
Population~III SFR contribution is larger in model 1. These results
are in good agreement with the hydrodynamic simulations of
\citet{2012ApJ...745...50W}, who use only very massive Population~III
stars.  We find that while the amplitude of the Population~III SFR is
sensitive to this effect, the lowest redshift with nonzero cosmic
Population~III SFR density is more dependent on the evolution of the
minimum mass of star-forming haloes, $M_\mathrm{min}$.  Further, the
small contribution of Population~III stars to the global SFR is
reflected in the fact that the effect of changing the Population~III
stellar IMF on the reionization and thermal history of the IGM is
negligible in our model.  As a result, the temperature evolution, as
well as the photoionization rate evolution is practically independent
of the Population~III IMF.

However, we find that although the effect of Population~III on the SFR
and the IGM reionization and thermal history is small, it has a
significant effect on the chemical properties of low mass galaxies.
This is seen in Figure \ref{plot:abr_z6} (left column), which shows
abundance ratios of various metal species as a function of halo mass
at $z=6$ for our three models.  We focus on [C/Fe], [O/Si], [Zn/Fe],
and [N/O].  It is seen that in haloes with mass greater than a few
times $10^9$ M$_\odot$, all four abundance ratios are constant, and
equal to the Population~II values.  Changing the Population~III IMF
has no effect on the abundance ratios in these haloes.  This is
because in these haloes, Population~III star formation has ceased at a
much earlier time and the subsequent Population~II star formation has
wiped out any chemical signature of Population~III star formation.
The constancy of the abundance ratios is due to the fact that for a
time-independent IMF, abundance ratios are equal to the ratio of
corresponding chemical yields \citep{1980FCPh....5..287T,
  2009nceg.book.....P}.  In low mass haloes, on the other hand,
chemical signatures of Population~III star formation have not yet been
wiped out.  As a result, abundance ratios in these haloes depend on
the Population~III IMF and are different in each of our Population~III
models.  Thus, the chemical signature of Population~III stars slowly
emerges as we move towards low halo masses.  Our model suggests that
this is probably the best hope of constraining the Population~III SFR
density on cosmic time scale and the Population~III IMF.

When we change the Population~III model, the abundance ratios in low
mass haloes change significantly.  For example, the [O/Si] ratio shows
a change of more than 1 dex when we change from model 2 to model 3.
This can be seen in Table \ref{table:yields}, which shows the
Population~III yields of O and Si in all three models.  Here, the
yield of species $i$ is defined as
\begin{equation}
y_i=\frac{\int dm\cdot\phi(m)\cdot m\cdot p_i(m)}{\int dm\cdot m\cdot\phi(m)},
\label{eqn:yields}
\end{equation}
where $p_i(m)$ is the fraction of initial stellar mass $m$ that is
converted to species $i$.  (Table \ref{table:yields} shows
$\log(y_i)$.) The ratio of oxygen yield to silicon yield is comparable
to solar in model 1.\footnote{For a given element X, we define the
  abundance relative to the solar value in the usual way, i.e.,
  [X/Y]$=\log_{10}(\mathrm{n_X}/\mathrm{n_Y})-\log_{10}(\mathrm{n_X}/\mathrm{n_Y})_\odot$,
  where $n_i$ is the number density of element $i$.  Solar
  photospheric abundance values are taken from
  \citet{1989GeCoA..53..197A}.}  However, it is much lower than solar
in model 3, and much higher than solar in model 2.  This is because in
model 3 high mass Population~III stars produce large amounts of
silicon during the O-burning process.  This is reflected in the [O/Si]
ratio shown in Figure \ref{plot:abr_z6}.  Similar considerations
explain the differences seen in other abundance ratios in low mass
haloes in the three models.  As we move towards lower redshift, we
would expect to see the abundance ratio values to move away from their
Population~III values towards the Population~II values even in small
mass haloes.  This is seen in Figure \ref{plot:abr_z2} (left column)
which shows the same abundance ratios as Figure \ref{plot:abr_z6} at
$z=2$.  The abundance ratio values in low mass haloes are much closer
to the Population~II values.

\subsection{Why do low mass galaxies retain an imprint of Population~III stars?}
\label{sec:why}

\begin{figure}
  \begin{center}
  \includegraphics[scale=0.5]{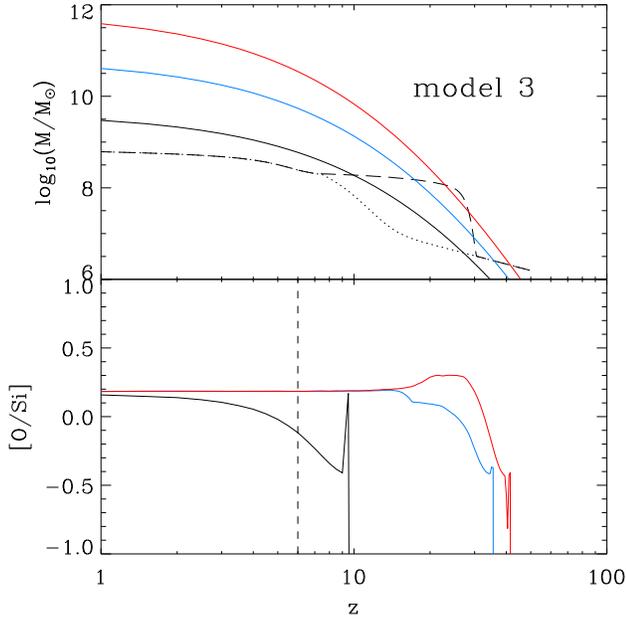}
  \end{center}
  \caption{Comparison of chemical evolution in haloes with three
    different masses, shown here for model 3.  The dashed line in the
    top panel shows the H~\textsc{ii} region filtering mass from
    Figure \ref{plot:mmin}, whereas the dotted line shows the
    evolution of the average filtering mass scale.  The black, red,
    and blue curves denote three different haloes.  The dashed
    vertical line marks $z=6$, which corresponds to the redshift in
    Figure \ref{plot:abr_z6}.  The spike in the plot is a result of
    the most massive stars in the model.} 
  \label{plot:sh}
\end{figure}

As described above, our model predicts that Population~III stars will
leave an imprint on the ratio of abundances of metals in low mass
haloes, while any such imprint is wiped out in high mass haloes by
subsequent Population~II star formation.  In this section, we discuss
the reason behind this contrast between low mass and high mass haloes.
In a nutshell, it is caused by a difference in Population~III-to-II
transition redshifts of these haloes, which in turn is a result of
photoionization feedback.

The solid curves in Figure \ref{plot:sh} (top panel) show the average
growth of three different haloes from H~\textsc{ii} regions in one of
our models (model 3).  These haloes have masses of $10^{9.5}$
M$_\odot$, $10^{10.5}$ M$_\odot$, and $10^{11.5}$ M$_\odot$
respectively at $z=1$.  Their growth is described by Equation
\ref{eqn:halo_assembly}.  The dashed line in this panel shows the
evolution of the filtering mass $M_\mathrm{min}$ in H~\textsc{ii}
regions in this model.  It can be seen that the three haloes satisfy
the star formation criterion, $M>M_\mathrm{min}$ at three different
redshifts, approximately $z=10$, $20$, and $25$ respectively.  Thus
each halo forms its first stars, which are Population~III, at very
different times.  This has an effect on the chemical evolution of
these three mass bins, which is shown in the bottom panel of Figure
\ref{plot:sh}.  In this panel, the three solid curves show the
evolution of [O/Si] in the same halo mass bins as the top panel.  (We
still focus on model 3.)  Since the Population~III IMF in model 3
(100--260 M$_\odot$) produces negative values of [O/Si], we expect a
negative value of [O/Si] in each halo mass bin when Population~III
stars form.  However, as time progresses, Population~II star formation
takes over in each mass bin, because of which the value of [O/Si]
approaches $+0.2$.  This is seen to happen in all three halo mass
bins.  However, this happens \emph{at different redshifts} for
different halo masses because they have different redshifts of
Population~III star formation (and therefore also the redshifts of
Population~III-to-II transition).  The redshift at which
Population~III star formation first occurs is the redshift at which
the condition $M>M_\mathrm{min}$ is first met, and is thus governed by
the thermal feedback.  As a result, at a given redshift, low mass
haloes retain memory of their Population~III enrichment history.  This
is highlighted by the vertical line marking $z=6$ on the bottom panel
of Figure \ref{plot:sh}.  It can be seen that at this redshift, the
two higher halo mass bins have already converged to [O/Si]$=0.2$, but
the lower mass bin still has [O/Si]$=-0.3$.  It is this effect that
causes the trends seen in Figure \ref{plot:abr_z2}.

\begin{figure}
  \begin{center}
  \includegraphics[scale=0.5]{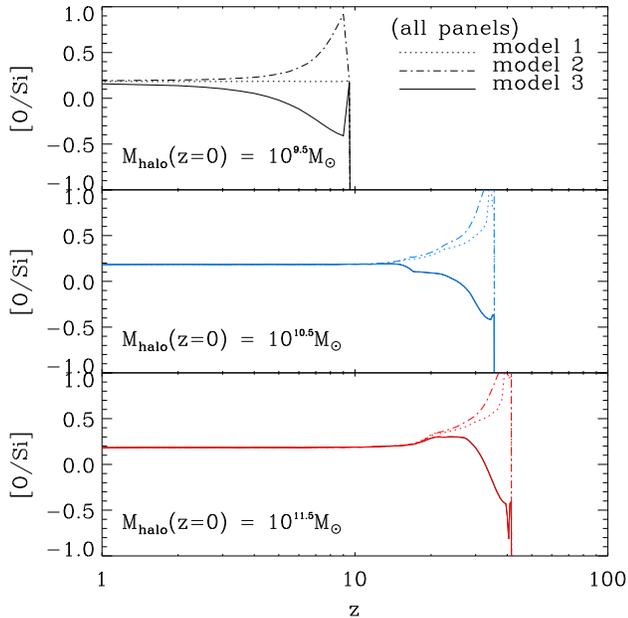}
  \end{center}
  \caption{Evolution of [O/Si] in our three models in the same three
    mass bins as in Figure \ref{plot:sh}, shown here in the same
    colours.  The dotted, dot-dashed, and solid curves show evolution
    in models 1, 2, and 3 respectively.}
  \label{plot:sh_obyfe}
\end{figure}

We now see why changing the Population~III stellar IMF will
selectively affect low mass haloes.  Figure \ref{plot:sh_obyfe} shows
the evolution of [O/Si] for the same three haloes as Figure
\ref{plot:sh} in each of our three Population~III models.  As expected
from the yields in Table \ref{table:yields}, [O/Si] for Population~III
stars in models 1 and 2 is greater than $0.2$, but is less that $0.0$
for model 3.  The curves for models 1 and 2 show an enhancement in
[O/Si] during the first star formation episode in each halo mass bin.
But this occurs at different redshifts for different halo masses, and
explains the variation in the trends seen in Figure \ref{plot:abr_z2}.
An alternate way of understanding this is that in a closed system with
a non-evolving IMF, all abundance ratios get locked at the value given
by the ratio of their respective yields \citep{1980FCPh....5..287T,
  2009nceg.book.....P}
\begin{equation}
  \frac{Z_\mathrm{O}}{Z_\mathrm{Si}}=\frac{y_\mathrm{O}}{y_\mathrm{Si}},
\end{equation}
where the yields are defined in Equation (\ref{eqn:yields}).  While
this is strictly valid only under the instantaneous recycling
approximation, it holds for real galaxies to a very good approximation
\citep{1980FCPh....5..287T}.  As a result, abundance ratios will
change only when there is a change in the IMF.  This is exactly what
happens when the stellar IMF in a galaxy changes from Population~III
to Population~II.  For low-mass galaxies, this happens at a lower
redshift.  Therefore, they retain the Population~III signatures at
these redshifts.  Note that the gas phase metallicity in these
galaxies is greater than $Z_\mathrm{crit}$, so they have already
stopped forming Population~III stars.  Still, their relative
abundances have not yet settled on the Population~II values, and
therefore carry a dependence on the Population~III IMF.

It is often mentioned in the literature that an ``odd-even'' pattern
in the abundances is a signature of Population~III stars.  This refers
to the fact that in metal-free stars the production of species with
odd nuclear charge is preferentially suppressed relative to their
solar values.  As we have seen above, the low mass haloes in our
models retain memory of Population~III star formation and therefore
also show the odd-even pattern in elemental abundances.  However, it
is important to understand that odd-even pattern is not unique to
Population~III and also occurs in metal-poor Population~II stars
\citep{2011MNRAS.417.1534C}.  Hence the crucial point here is that one
really wants to search for relative abundance trends which differ from
the Population~II predictions, and this information is encoded in
low-mass halos.

\subsection{Role of critical metallicity}
\label{sec:zcrit}

In all of above, the critical metallicity $Z_\mathrm{crit}$, at which
a Population~III IMF changes over to a Population~II IMF, plays a
crucial role.  However, the value of $Z_\mathrm{crit}$ is currently a
topic of debate \citep{2012MNRAS.421.3217K, 2012arXiv1203.6842D}.
When metallicity $Z$ is greater than $Z_\mathrm{crit}$, the
availability of metal-line cooling and/or dust cooling is expected to
lead to fragmentation in the ISM that could help form small mass stars
with a Population~II type IMF.  In this paper, we have assumed
$Z_\mathrm{crit}=10^{-4}Z_\odot$, a value motivated by most studies
of fragmentation in metal-poor gas \citep{2001MNRAS.328..969B,
  2003Natur.425..812B, 2007MNRAS.380L..40F}.  However, some studies
have argued for a much smaller value of
$Z_\mathrm{crit}=10^{-6}Z_\odot$ \citep{2012MNRAS.421.3217K}.  Since
this number is not very well constrained, we comment on the effect on
our results of variation in its value.

We find that values of $Z_\mathrm{crit}$ smaller than $10^{-4}Z_\odot$
have a very minor influence on our results.  This is because in all
three of our models, the metallicity of ISM in all haloes crosses
$Z_\mathrm{crit}$ in a very short time after the first burst of
Population~III star formation.  We explicitly checked this in our
calculation by changing $Z_\mathrm{crit}$ to $10^{-5}Z_\odot$ and
$10^{-6}Z_\odot$.  In our models 2 and 3, in which the Population~III
metal yield is higher, there is no effect of a change in the critical
metallicity.  In model 1, lowering the critical metallicity to
$10^{-6}Z_\odot$ has an effect of decreasing the average
Population~III SFR density, but only for $z>20$.  When $Z_\mathrm{crit}$
is lowered, haloes in this model take shorter time to stop forming
Population~III stars.  These results are in qualitative agreement with
those of \citet{2007MNRAS.381..647S}, who also found that the
influence of $Z_\mathrm{crit}$ depends on the Population~III stellar
IMF.  Given the negligible effect of changing $Z_\mathrm{crit}$, we
keep $Z_\mathrm{crit}=10^{-4}Z_\odot$ in this paper.

\subsection{Observability as DLAs}

\begin{figure}
  \begin{center}
  \includegraphics[scale=0.5]{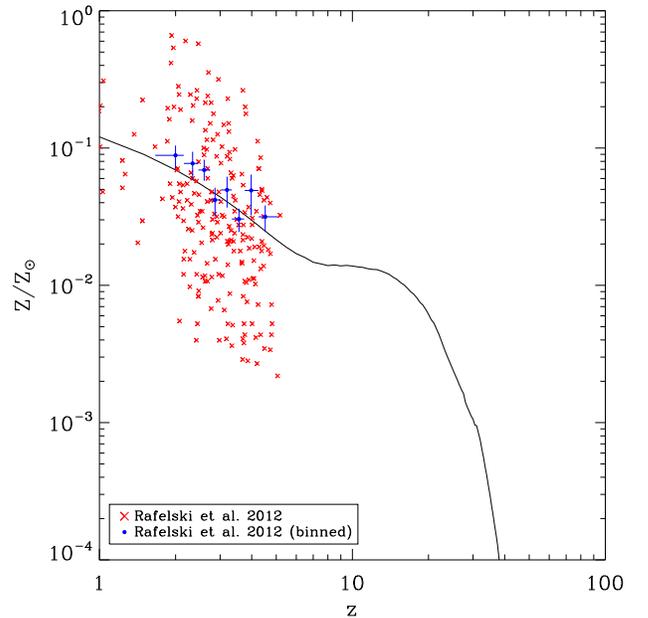}
  \end{center}
  \caption{The ``age-metallicity'' relation in our model, which is
    given by an absorption cross section weighted average of gas phase
    metallicity of all mass bins.  Red crosses are measurements of a
    sample of 241 DLAs by \citet{2012ApJ...755...89R}.  The blue
    points are mean values in different redshift bins, as presented by
    \citet{2012ApJ...755...89R}.  Vertical error bars on the blue
    points represent 1$\sigma$ error bars.}
  \label{plot:zism}
\end{figure}

As we saw above, low mass galaxies (halo masses $\sim 10^9$ M$_\odot$)
at high redshift ($z\sim 6$) are likely to be useful in constraining
Population~III IMF and SFR.  However, these galaxies are difficult to
observe.  Even otherwise, measurements of their relative abundances
are unlikely to be accurate, as measurements from line emissions
involve highly uncertain modeling of the observed H~\textsc{ii}
regions.  As a result, we need to look for an alternative method of
observing these galaxies and measuring their metal abundances.  We now
argue that damped Ly$\alpha$ systems (DLAs) in high-redshift quasar
spectra provide just such an alternative.

DLAs represent H~\textsc{i} reservoirs at high redshift.  Metal
abundances of DLAs have been measured up to $z=6$.  Using this data,
an ``age-metallicity relation'' can be constructed for DLAs.
Moreover, chemical abundance measurements in DLAs are very accurate.
This is because in the cold H~\textsc{i} reservoirs of DLAs, most
metal species are in their singly ionised state, which can be probed
in the optical from ground-based telescopes.  Errors in DLA
H~\textsc{i} column density measurements are typically quite low,
around 0.05 dex.  Therefore errors in the corresponding metal
abundance measurements ([M/H]) is also low, around 0.1 dex.  Thus,
observations of DLAs can be used to measure gas-phase metallicities at
large cosmological lookback times with high precision
\citep{2005ARA&A..43..861W}.  Furthermore, in DLAs, \emph{relative
  abundances} can still be measured accurately deep into the
reionization epoch ($z > 6$) using metal-line transitions redward of
Ly$\alpha$, even though Gunn-Peterson absorption precludes measurement
of neutral hydrogen \citep[cf.][]{2012ApJ...744...91B}.

In order to predict the properties of DLAs in our model, we assign a
mass-dependent, physical, neutral hydrogen cross-section (``size'') to
each halo, at every redshift.  This assignment is performed using the
fitting function
\begin{equation}
\Sigma(M)=\Sigma_0\left(\frac{M}{M_0}\right)^2\left(1+\frac{M}{M_0}\right)^{\alpha-2},  
\label{eqn:dlafit}
\end{equation}
where the constants take the values of $\alpha=0.2$, $M_0=10^{9.5}$
M$_\odot$, and $\Sigma_0=40$ kpc$^2$ at $z=3$
\citep{2008MNRAS.390.1349P, 2012arXiv1209.4596F}.  Values at other
redshifts are calculated by mapping haloes at these redshifts to
haloes $z=3$ according to circular velocity
\citep{2012arXiv1209.4596F}.  Our choice of the fitting function in
Equation (\ref{eqn:dlafit}) is inspired by
\citet{2008MNRAS.390.1349P}, who find that it provides a good fit to
absorption systems in their hydrodynamical simulation at $z=3$.  As
another check on Equation (\ref{eqn:dlafit}), we look at the average
evolution of ISM metallicities.  The average gas-phase metallicity at
a given redshift is
\begin{equation}
  Z(z) = \frac{\int dm\cdot N(m,z)\cdot Z_\mathrm{ISM}(m,z)\cdot
    \Sigma(m,z)}{\int dm\cdot N(m,z)\cdot \Sigma(m,z)}.
\end{equation}
Figure \ref{plot:zism} shows this quantity in comparison with
metallicity estimates for 100 damped Ly$\alpha$ systems (DLAs) with
$z\lesssim 4$ by \citet{2003ApJ...595L...9P}.  Our models are in good
agreement with the mean metallicities of the observed DLAs.  It is
encouraging that our results are also comparable to those of the
recent hydrodynamical simulations \citep{2012ApJ...748..121C,
  2011MNRAS.418.1796F, 2011MNRAS.415...11D}.

\begin{figure}
  \begin{center}
  \includegraphics[scale=0.5]{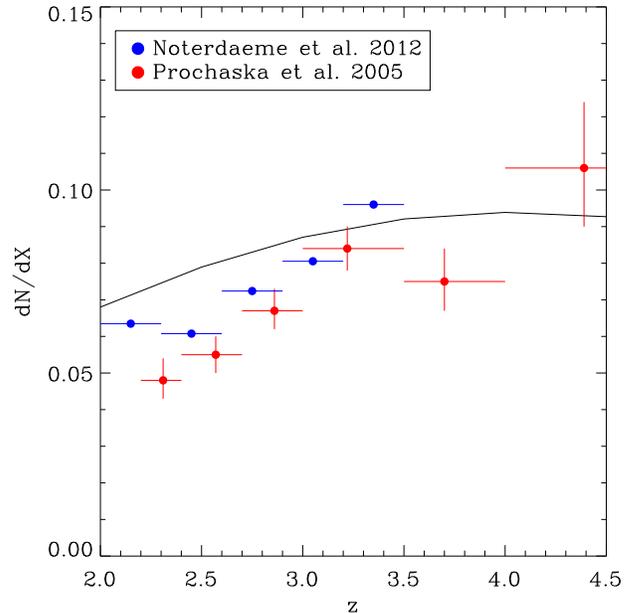}
  \end{center}
  \caption{Evolution of $dN/dX$ in our model.  The blue and red data
    points are from measurements by \citet{2012A&A...547L...1N} and
    \citet{2005ApJ...635..123P} respectively.}
  \label{plot:ldla}
\end{figure}

For any measurable property $p$ (such as metallicity, or abundance
ratio) of DLAs, we can calculate the number of systems with different
values of $p$ in a sample of DLAs.  This is called the line density
distribution, and with Equation (\ref{eqn:dlafit}) in hand, it can be
written as \citep[e.g.,][]{2005ARA&A..43..861W}
\begin{equation}
\frac{d^2N}{dXdp}=N(M)\cdot\Sigma(M)\cdot\frac{dl}{dX}\frac{dM}{dp}\cdot (1+z)^3.
\label{eqn:d2n}
\end{equation}
Here, $X$ is an absorption length element given by
\begin{equation}
\frac{dl}{dX}=\frac{c}{H_0(1+z)^3},
\label{eqn:dx}
\end{equation}
$dl=cdt$ is a length element, and $p$ is the property in
consideration.  The halo mass is denoted by $M$, $N(M)$ is the
comoving number density of halos (i.e., the halo mass function), and
$\Sigma(M)$ is the halo cross section given by Equation
(\ref{eqn:dlafit}).  The quantity $dM/dp$ in Equation (\ref{eqn:d2n})
can be easily calculated in our model, as properties like metallicity
and relative abundances are known for all halo masses.  The integral
of Equation (\ref{eqn:d2n}) over all values of $p$ is just the total
line density of DLAs, $dN/dX$.  The evolution of $dN/dX$ in our model
is shown in Figure \ref{plot:ldla}.  Our results are in reasonable
agreement with the values obtained from the Sloan Digital Sky Survey
(SDSS) DR5 by \citet{2005ApJ...635..123P} and from SDSS DR9 by
\citet{2012A&A...547L...1N}. (Note that we have assumed a slightly
smaller value of $\alpha$ compared to \citet{2008MNRAS.390.1349P} to
get a good match to observed $dN/dX$.)

Using Equation (\ref{eqn:d2n}), we plot the line density distribution
of DLAs at $z=6$ as a function of various metal abundance ratios in
Figure \ref{plot:abr_z6} (right column).  We focus on [O/Si] for this
discussion, which can be easily generalized for other ratios in Figure
\ref{plot:abr_z6}.  The panel corresponding to [O/Si] shows the DLA
line density distribution for the three Population~III IMFs considered
in this paper (Table \ref{table:models}).  The distribution
corresponding to model 2 (35--100 M$_\odot$; blue curve) is peaked at
[O/Si]$=0.2$ and has a tail spreading out to higher values up to
[O/Si]$=0.8$.  The distribution corresponding to model 3 (100--260
M$_\odot$; red curve) is also peaked at [O/Si]$=0.2$, but has a tail
spreading out to \emph{lower} values down to [O/Si]$=-0.4$.  Finally,
model 1 (1--100 M$_\odot$) is simple a delta function at [O/Si]$=0.2$.
The line density distribution in this model has no tail; all DLAs have
the same abundance ratio of [O/Si]$=0.2$.  This suggests that the
distribution of relative abundance values in a sample of DLAs depends
on the Population~III IMF, at least at sufficiently high redshifts.

This dependence of the line density distribution on Population~III IMF
can be understood in terms of the dependence of halo relative
abundances on the Population~III IMF.  We discussed the latter in
sections \ref{sec:pop3effect} and \ref{sec:why} above, and in the
left-hand column of Figure \ref{plot:abr_z6}.  Focusing again on
[O/Si], recall that most high-mass haloes had [O/Si]$=0.2$, while
low-mass haloes had different values of [O/Si], depending on the
Population~III IMF.  This same effect is reflected in the line density
distribution in the right-hand column of Figure \ref{plot:abr_z6},
since all of these haloes contribute to the line density distribution
via Equation (\ref{eqn:d2n}).  Thus, e.g., as low-mass haloes in model
2 (35--100 M$_\odot$; blue curve) have [O/Si]$>0.2$, DLAs
corresponding to these haloes form the tail that spreads towards this
values.  Similarly, in model 3 (100--260 M$_\odot$; red curve),
low-mass haloes have negative [O/Si], which results in a tail in the
DLA line density distribution that extends towards these values.
Finally, in model 1 (1--100 M$_\odot$), all haloes have the same value
of [O/Si].  As a result, the DLA line density is a delta function
centered on this value (0.2) and has no tail.  This also suggests that
with decreasing redshift, as more and more haloes move to their
Population~II-producing phase, the DLA line density distribution will
reduce its spread and move towards the Population~II value.  This is
exactly what is seen as we move from $z=6$ to $z=2$, as shown in
Figure \ref{plot:abr_z2} (right column).

In sum, Figure \ref{plot:abr_z6} displays the general result that
\emph{the distribution of DLAs in the abundance-ratio space at
  sufficiently high redshift is sensitive to the Population~III IMF}.
In the absence of Population~III stars causing any change in stellar
yields over time, this distribution will be a delta function in the
abundance-ratio space (neglecting corrections for the impact of dust
on relative abundances, which we discuss below).  However, the
addition of a new Population~III stellar population with different
chemical yields, spreads DLAs out in the abundance-ratio space.  The
shape of this spread can constrain the IMF of Population~III stars,
while its evolution can constrain the Population~III SFR history.
(Note that as discussed before in section \ref{sec:why} above, the gas
phase metallicity of these DLAs is higher than $Z_\mathrm{crit}$,
similar to a vast majority of observed DLAs.  Still their relative
abundances have not yet settled on the Population~II values.  This
results in a dependence on the Population~III IMF.)

\begin{figure}
  \begin{center}
  \includegraphics[scale=0.5]{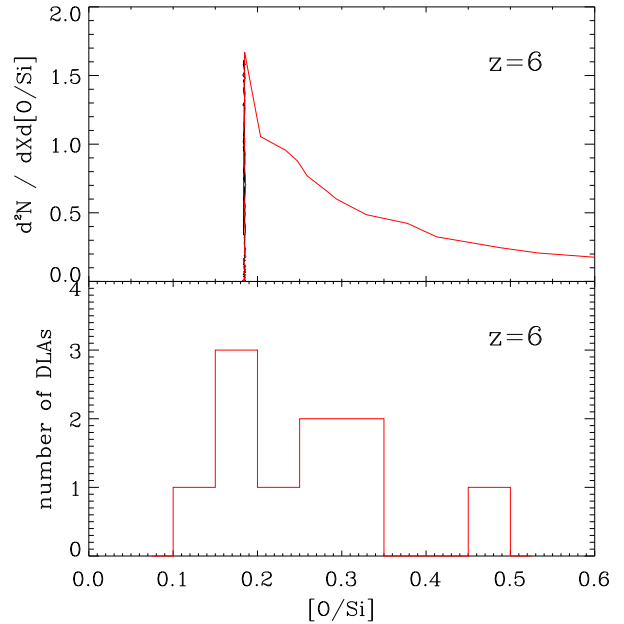}
  \end{center}
  \caption{Top panel shows the predicted distribution of [O/Si] values
    for Population~II (black) and Population~III (red; model 2) at
    $z=6$.  (This distribution is taken from Figure
    \ref{plot:abr_z6}.)  The bottom panel shows a histogram of a
    random sample of 10 [O/Si] measurements taken from the
    Population~III distribution in the top panel.  An additional
    Gaussian error of 0.1 dex is added to each sampled point.  A KS
    test shows that this sample rejects the Population~II distribution
    at approximately 4$\sigma$ ($D=0.9$, $p=0.00017$).  This suggests
    that it is possible to detect the effect of Population~III star
    formation even with a relatively small sample of DLA abundance
    ratio measurements with an accuracy of about 0.1 dex.}
  \label{plot:sample}
\end{figure}

Further, to check whether this effect is detectable and to show that
it can indeed be used to probe the Population~III IMF, in Figure
\ref{plot:sample}, we show a simulated data set of 10 [O/Si]
measurements at $z=6$ taken from the DLA line density distribution
corresponding to model 2 (35--100 M$_\odot$; shown in the top panel of
Figure \ref{plot:sample} and also in Figure \ref{plot:abr_z6}).  This
data set is produced by random sampling the predicted distribution of
[O/Si] values in the model, and adding a Gaussian error of 0.1 dex to
each sample point.  An error of 0.1 dex corresponds to the typical
accuracy with which abundance ratios in DLAs are measured
\citep[e.g.,][]{2012ApJ...744...91B}.  Using a KS test, we find that
this sample rejects the Population~II distribution at 4$\sigma$
($D=0.9$, $p=0.00017$).  A set of 100 samples (each of size 10 and a
Gaussian error of 0.1 dex) preferred the Population~III IMF at
3.8$\sigma$ on average ($\langle D\rangle=0.8, \langle
p\rangle=0.007$).  This suggests that the effects of Population~III
IMF on DLA abundance ratio are significant enough to be detectable
with just 10 measurements accurate to about 0.1 dex\footnote{Any
  effects of dust depletion could in principle be higher than 0.1 dex.
  However, the scatter in observations at low redshift ($z\sim 2$) is
  less than 0.2 dex \citep{2012ApJ...744...91B}, which suggests that
  our results are valid even when effects of dust are included.}.  A
larger sample of 20 measurements rejects the Population~II
distribution at an even higher significance of about 5$\sigma$
($\langle D\rangle=0.85, \langle p\rangle=10^{-5}$; again assuming an
accuracy of 0.1 dex).  Note that this test used a single abundance
ratio ([O/Si]).  In practice, using multiple ratios as shown in Figure
\ref{plot:abr_z6} can further improve the significance of the
constraints.  Also note that the metallicity of these low mass halos
($\mathrm{M}\sim 10^9 \mathrm{M}_\odot$) with Population~III
signatures is high enough to produce detectable lines at $z\sim 6$.
For example, we can estimate the O~\textsc{i} column density by
\citep{2012MNRAS.421L..29S}
\begin{equation}
  N_\mathrm{OI}=\frac{3}{2\pi}\frac{M_\mathrm{O}/\mu_\mathrm{O} m_p}{\alpha^2r_\mathrm{vir}^2},
\end{equation}
where $\mu_\mathrm{O}$ is the atomic weight of oxygen and $r=\alpha
r_\mathrm{vir}$ is the gas radius, where usually $\alpha\sim 0.8$
\citep{2012MNRAS.421L..29S}.  The oxygen mass $M_\mathrm{O}$ is
calculated in each halo in our chemical evolution model.  This gives
$N_\mathrm{OI}>10^{15}$ cm$^{-2}$ for $\alpha=0.8$.  This value of the
column density is already higher than that measured in several $z=6$
DLAs by \citet{2012ApJ...744...91B}, which span the range
$N_\mathrm{OI}=10^{13.49}$--$10^{14.47}$ cm$^{-2}$.

\begin{figure}
  \begin{center}
  \includegraphics[scale=0.5]{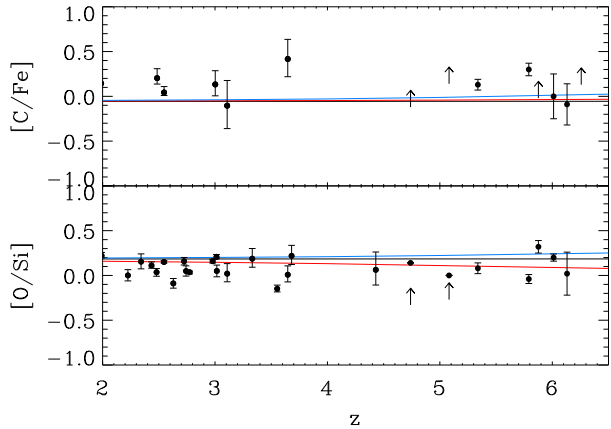}
  \end{center}
  \caption{Evolution of the \emph{average} values of [C/Fe] and [O/Si]
    in our models.  The black, blue, and red curves correspond to
    models 1, 2, and 3 respectively.  Data points show relative
    abundances in DLAs from a compilation by
    \citet{2012ApJ...744...91B}.  It is seen that changing the
    Population~III IMF has little effect on the average values of
    relative abundances.  However, the effect on the distribution of
    abundance ratio values is significant and detectable, as shown in
    Figure \ref{plot:sample}.  This shows that it is the distribution
    of measurements that probes the stellar IMF and not the mean value
    of the measurements.}
  \label{plot:avg_cbyfe}
\end{figure}

It is important to understand that the \emph{distribution} of relative
abundance values at high redshifts is used here, not the mean value at
those redshifts.  In fact, as we have seen, all line density
distributions in Figure \ref{plot:abr_z6} are anchored on the
Population~II values of the respective abundance ratios.  This is
because even at high redshift, high mass haloes have already moved to
their Population~II star forming stage; only low mass haloes are
expected to carry the memory of their Population~III enrichment.
Therefore, we expect the mean value of these any abundance ratio to be
close to the Population~II value.  It will have very little dependence
on the Population~III IMF.  This can be understood by explicitly
calculating the mean value of different abundance ratios, given by
\begin{equation}
  \langle p\rangle=\int\frac{d^2N}{dXdp}\cdot p\cdot dp\cdot dX, 
\end{equation}
where $p=$[M$_1$/M$_2$] is the abundance ratio of two species M$_1$
and M$_2$, and $dX$ is defined in Equation (\ref{eqn:dx}).  Figure
\ref{plot:avg_cbyfe} shows the mean value of [C/Fe] and [O/Si]
calculated in this fashion, corresponding to our three Population~III
models.  Observational measurements of several DLAs from
\citet{2012ApJ...744...91B} are also shown.  It is seen that the
average values predicted by our model are consistent with the
observations, regardless of the Population~III IMF used.  (Note that
the observational data shown here is uncorrected for effects of dust
depletion.)  The most striking feature in Figure \ref{plot:avg_cbyfe}
is the lack of evolution over the large redshift range from $z\gtrsim
6$ to $z\sim 2$.  This is explained in our model by the fact that all
large haloes show constant abundance ratios, corresponding to
Population~II.  A second feature of Figure \ref{plot:avg_cbyfe} is
that varying the Population~III IMF results in only a small change in
the average value of the abundance ratio, as expected.  It is only
when we look at the spread of these values at different redshift that
the effect of Population~III IMFs becomes visible.

\section{Discussion}
\label{sec:discuss}

We have shown that a chemical evolution model that also incorporates
global effects like reionization and photoionization feedback is
consistent with a wide range of possible Population~III IMFs.
However, this approach uncovers patterns in the metallicity
distribution of high redshift galaxies which could potentially be
useful in ruling out some Population~III IMFs and SFRs.

It is therefore important to understand the domain of validity of our
model.  \citet{2004Ap&SS.294...29M} and \citet{2009A&A...504..373C}
have studied the evolution of the mass-metallicity relation and the
chemical properties of systems like dwarf galaxies using chemical
evolution modelling of the kind that we have used here.  They find
that if the instantaneous recycling approximation is avoided, an
approach that correctly incorporates effects of outflows and inflows
successfully reproduces all observed chemical properties of galaxies.
\citet{2004MNRAS.349.1101D} and \citet{ 2012MNRAS.426L..61D} have also
used a similar chemical evolution model in their semi-analytic model
of galaxy formation that fits several key observations very well.
Furthermore, we focus on abundance ratios in this paper, which are
more robustly predictable, given their simple dependance on the
stellar IMF.  Thus, given the uncertainties in high redshift
observations and the simplicity of our model, the main conclusion of
this paper regarding chemical signatures of Population III stars in
high redshift galaxies is robust.

A second uncertainty in our calculation is the prescription used for
photoionization feedback.  Firstly, we have used a simple Jeans
prescription whereas it has been argued that more detailed
prescriptions give a better fit to hydrodynamical simulations
\citep{2000ApJ...542..535G, 2008MNRAS.390..920O}.  However, the
difference between these feedback prescriptions is small, especially
at $z\gtrsim 5$ \citep{2001PhR...349..125B}.  A detailed understanding
of radiative feedback in galaxy formation is yet to emerge, primarily
because of the large variety of approximations used in earlier work
\citep{2005SSRv..116..625C}.  However the general consensus is that
such a filtering scale increases from roughly $10^7$ M$_\odot$ at
$z=10$ to $10^9$ M$_\odot$ at redshift $z=6$.  This is in good
agreement with the values we obtain in our study.  A tentative
observational evidence of such effect is seen in ``dark'' dwarf
galaxies in the vicinity of high redshift AGN
\citep{2012MNRAS.425.1992C} and in the difference between the
evolution of star formation rate in low mass galaxies and that in high
mass galaxies \citep{2012MNRAS.426.2797W}.  Some authors have used
simple fitting functions for $M_\mathrm{min}$ that agree quite well
with our evolution of $M_\mathrm{min}$ \citep{2012MNRAS.421L..29S}. A
similar approach has also been used in semi-analytical models of
galaxy formation \citep[e.g.,][]{2006RPPh...69.3101B}.  It should also
be noted that the evolution of $M_\mathrm{min}$ for $z\lesssim 5$ is
not important for our present study as by such low redshift the
contribution of Population~III stars to chemical evolution is
negligible.

Thirdly, our approach of evolving haloes and galaxies \emph{in the
  mean} does not provide us with information regarding the scatter
around the distributions shown here.  However, we note that
semi-analytic models based on halo merger trees find quite small
scatter around the mean distributions of metallicities \citep[around 1
  dex at redshift 0;][]{2004MNRAS.349.1101D}.  Further, scatter does
not affect abundance ratios; it can only affect the star formation
histories of individual haloes.  Therefore the effect of scatter on
our central prediction, that distribution of DLA abundance ratios
probe Population~III IMFs, will be nearly zero.  For example, positive
values of [O/Si] at high redshift are inconsistent with Population~III
stars in the PISN range.  Observations of these systems would be a
strong argument against high mass Population III stars, regardless of
the scatter in the observations, which will not affect the [O/Si]
ratio.  In fact, we want to highlight the fact that in this respect
this approach of constraining Population III scenarios from
observations of high redshift DLAs is possibly better than
constraining them using observations of metal-poor stars in the
Galactic halo.  The metallicity distribution of metal-poor stars is
highly dependent on the peculiar assembly history of the Milky Way,
which is not guaranteed to follow the ``average'' assembly history
that models require in order to constrain Population~III SFR and IMF.
As such, metal-poor stars are useful only if they unambiguously show
the signature of a pair instability supernova.  However, even in this
case they will only provide indication of a single Population~III star
of a certain mass, not of the whole IMF or SFR.\footnote{Another way
  in which metal-poor stars can be of use is if we find a truly
  metal-free star.  But this is difficult given the observational
  limitations \citep{2011arXiv1102.1748F}.}  Similar arguments hold
for observations of dwarf spheroidals and ultra-faint dwarfs.

We have assumed a constant UV photon escape fraction while calculating
the IGM reionization history.  This can potentially affect our
conclusions if the escape fraction in Population~III star forming
haloes is so large that reionization constraints begin restricting
Population~III star formation parameter space.  Indeed, it has been
argued that the escape fraction will have a strong increasing
evolution with redshift \citep{2012MNRAS.423..862K,
  2012MNRAS.tmpL...1M}.  Also, the escape fraction could potentially
have different values for galaxies with Population~II and
Population~III star formation due to different spectral indices
\citep{2006MNRAS.371L..55C}.  However, our assumption about the escape
fraction is unlikely to affect our results significantly since we find
that Population~III stars do not contribute much to reionization due
to their early termination.  Thus, the reionization history affects
our conclusions only moderately.  Note that the value of escape
fraction used in our model is comparable to that deduced from
luminosity function measurements in recent studies
\citep{2012MNRAS.tmpL...1M}.

Finally, we comment about the role of dust.  Dust affects the metal
evolution of galaxies by preferentially depleting certain species like
iron.  We have not taken this effect into account.  As such our
results are only ``production-side'' estimates of the chemical
abundance patterns.  Dust obscuration is ignored in most DLA studies
\citep{2005ARA&A..43..861W}.  This difficult problem can be alleviated
by focusing on non-refractory species such as zinc and oxygen, which
are not depleted on dust grains. We can also focus on ``secondary
elements'' like nitrogen that depend on the square of the metallicity,
which increases their sensitivity to underlying chemical enrichment
patterns.  We defer a detailed study of the effect of dust depletion
to a future work.  There is however, an empirical reason why dust
depletion may have very little effect on our main result.  This is
because it has been noted that dust depletion is strongly dependent on
metallicity such that low metallicity systems have very little dust
\citep{2004A&A...421..479V, 2012ApJ...744...91B, 2012ApJ...755...89R}.
This combined with the observed metallicity evolution shown in Figure
\ref{plot:zism} suggests that there is very little dust in DLA at
$z\sim 5$.  Furthermore, \citet{2012ApJ...755...89R} report a sudden
decrease in DLA metallicity at $z>4.7$ (at low significance), which if
real would further reduce the role of dust in these systems.  

\section{Conclusions}
\label{sec:conclude}

Our results are as follows: 

\begin{itemize}
\item We have developed a chemical evolution model of galaxies that is
  consistent with a variety of global observational constraints on
  galaxy and IGM evolution, such as the cosmic SFR density evolution,
  IGM thermal evolution, and hydrogen photoionization rate evolution.
  We calculate the minimum mass of star forming galaxies
  self-consistently.  This model produces galaxies that lie on
  observational curves such as the stellar-to-halo mass relation at
  low redshift and the mass-metallicity relation.  We then explore
  influence of Population~III stars on the predictions of this model,
  by varying the Population~III stellar IMF.
\item We find that different Population~III stellar IMFs result in
  very different abundance ratio distributions in low mass galaxies
  (halo mass $\lesssim 10^9$ M$_{\odot}$) at high redshift. This is
  because photoionization feedback suppresses star formation in these
  galaxies till low enough redshift ($z\sim 10$), and memory of the
  initial generation of Population~III stars is retained.  For some
  ratios, e.g., [O/Si], the variation is as large as 1 dex.  This
  variation is much greater than the uncertainty in metallicity
  measurements in DLAs.  This effect is strong at redshift $z\gtrsim
  5$ and grows weaker at lower redshift due to subsequent
  Population~II star formation.
\item The influence of Population~III star formation is seen only in
  low-mass haloes.  Changing the Population~III IMF has no effect on
  the abundance ratio in high mass haloes.  This is as expected since
  in these haloes abundance ratios are determined by Population~II
  stars alone.  It is only as we move downwards in halo mass that the
  Population~III signature starts dominating.
\item We modelled low-mass haloes as DLAs by assigning them a
  mass-dependent H~\textsc{i} absorption cross-section.  This model
  agrees with the observed metallicity-redshift relation for DLAs and
  the observed DLA line density evolution as a function of redshift.
  Using this model, we predict the line density distributions of DLAs
  as a function of different abundance ratios.  We find that these
  distributions are anchored towards the Population~II values of
  abundance ratios, but they show a significant dependence on the
  Population~III IMF for $z\gtrsim 5$.  This dependance is reduced at
  lower redshift due to subsequent Population~II star formation.
\item The dependance of the DLA line density distribution on
  Population~III suggests that the distribution of DLAs in
  abundance-ratio space at sufficiently high redshift can provide very
  good constrains on Population~III properties.  The form of this
  distribution can constrain the IMF of Population~III stars, while
  its evolution can constrain the global Population~III SFR history.
  This constraint on Population~III could possibly be stronger than
  constraints from other probes such as metal-poor stars and
  individual metal-poor DLAs.  A simulated data set of just 10 DLAs at
  $z\sim 6$ measured with realistic accuracy is able to constrain
  specific Population~III IMFs with high confidence.  This method of
  probing Population~III stars does not rely on the measurement of the
  H~\textsc{i} column density.  It is therefore useful at high
  redshift where Gunn-Peterson absorption precludes measurement of
  neutral hydrogen.  Relative abundances can still be measured
  accurately in this epoch using transitions redward of Ly$\alpha$,
  and self-schielded systems can be selected based on the presence of
  O~\textsc{i}.
\item Not only can the abundance ratio distributions distinguish
  between high-mass and low-mass Population~III IMFs, they can also
  discriminate between different low mass Population~III IMFs.  For
  example, the two low mass Pop~III IMFs that we consider in this
  paper are clearly distinguished by their oxygen abundance patterns.
\end{itemize}

\section*{Acknowledgements}

We would like to acknowledge useful discussions with Carlton Baugh,
T.~Roy Choudhury, Gabriella De Lucia, Andrea Macci\`o, J.~Xavier
Prochaska, Marc Rafelski, Hans-Walter Rix, Stefania Salvadori,
R.~Srianand, and members of the ENIGMA group at MPIA, especially
G\'abor Worseck and Neil Crighton.  We also thank the anonymous
referee for several constructive comments that improved our
presentation.  GK would also like to thank Institut d'Astrophysique de
Paris for hospitality.  This work was partly supported by the French
Agence Nationale pour la Recherche (ANR) within the Investissements
d'Avenir programme under reference ANR-11-IDEX-0004-02 and via the
grant VACOUL (ANR-2010-BLAN-0510-01).

\bibliography{refs}
\end{document}